\documentclass[%
 reprint,
 amsmath,amssymb,
 aps,
prb
]{revtex4-2}
\usepackage{graphicx}
\usepackage{dcolumn}
\usepackage{bm}
\usepackage{hyperref}
\usepackage{booktabs}
\usepackage{gensymb}
\usepackage[utf8x]{inputenc}                          
\usepackage[english]{babel}
\usepackage{float}
\begin{document}

\title{%
       Cryogenic spin 3/2 nuclear quadrupole resonance: Spin relaxation and electric field gradient via Rabi frequency goniometry
       }
    \author{Ritik R.\ Modi}
    \email{rmodi3@gmu.edu}
    \affiliation{%
        Department of Physics and Astronomy, George Mason University,
        Fairfax, VA 22030, USA%
    }
    \author{Karen L.\ Sauer}
    \email{ksauer1@gmu.edu}
    \affiliation{%
        Department of Physics and Astronomy, George Mason University,
        Fairfax, VA 22030, USA%
    }
    \affiliation{%
        Quantum Science and Engineering Center, George Mason University,
        Fairfax, VA 22030, USA%
    }

\date{\today}
\begin{abstract}

 A straightforward way to determine the electric field gradient $-$ principal axes frame (EFG-PAF) on single crystal samples with spin 3/2 nuclei is demonstrated. Nuclear quadrupole resonance (NQR) spectroscopy is used to determine the Rabi frequency for $^{35}$Cl in a single crystal of potassium chlorate (KClO$_3$) by comparing the NQR signal for powder and single crystal samples. By exploiting the geometrical dependence of the Rabi frequency with respect to the excitation direction, EFG-PAF is readily determined. Furthermore, relaxation times $T_1$ and $T_2^*$ were measured at multiple temperatures ranging from $17~\textrm{K}$ to $200 ~$K, extending the results of previous works to colder temperatures where new relaxation mechanisms become dominant. The experiments were performed in a cryogen-free cryostat, which posed distinct challenges compared to a conventional cryogenic cooling setup. The successful operation of the NQR probe within a cryogen-free cryostat has the potential to make the technique more accessible and widen applications.
 
\end{abstract}

\maketitle

\section{Introduction}

Studying the electric field gadient (EFG) provides crucial information about the local electrical environment of the nucleus. Thus, it has long been studied to understand the intrinsic properties of materials with nuclei having spins greater than $1/2.$ The EFG depends on the crystal or electronic structure \cite{grant1969,colville1970,iglesias2001,vojvodin2022,hartman2024}, temperature \cite{christiansen1976,nishiyama1976,jena1981,haas2024,laguta2023}, chemical bonds \cite{dejong1998,herzig2008,polak2003}, crystal defects \cite{ansari2023, sauer2003, suits2006, hughes1993, kanert1969, cohen1957, han1988, bowers2005, lany1980}, and more \cite{ansari2024,fenta2021,fujii2024,bonfa2022,rodrigues2020,song2022,banerjee2021}. Therefore, it would be useful to know the value and the orientation of the EFG. Nuclear quadrupole resonance (NQR) spectroscopy is a technique often used to study the EFG. A general overview of NQR spectroscopy can be found in Ref. \cite{suits2006}, and Ref. \cite{das1958}.\\

NQR uses radio-frequency (RF) excitation pulses to induce transitions between the quantum levels of the nucleus, similar to nuclear magnetic resonance (NMR) spectroscopy. However, unlike NMR, where the splitting is achieved by an external static magnetic field, the splitting in NQR arises due to the interaction between the quadrupolar nucleus and the intrinsic EFG. One of the goals of an NQR measurement is to determine the quadrupole coupling constant ($\nu_Q = eV_{zz}Q$), and the asymmetry parameter ($\eta  = \frac{V_{xx}- V_{yy}}{V_{zz}}$) \cite{suits2006}, where, $e$ is the elementary charge, $Q$ is the nuclear quadrupole moment \cite{abragam1961}, $V_{ij}$ is the EFG tensor with $x,y, \textrm{ and } z$ as its principal axes, and it can be defined as a real, symmetric, and traceless tensor, i.e., $\sum_{i=x,y,z} V_{ii} = 0$ \cite{suits2006}. It is customary to choose the principal axes such that $V_{zz} \ge V_{yy} \ge V_{xx}$ so that $0 \le \eta \le 1$.\\

However, for spin $3/2$ nuclei, there are two doubly degenerate energy levels, and only one (nonzero) transition frequency exists. This implies that $\eta$ and $\nu_Q$ cannot be determined by a conventional NQR measurement \cite{suits2006}. To break this degeneracy in the magnetic quantum number `$m$,' a small static magnetic field can be supplied \cite{ansari2024}, and a goniometer setup is then used to determine the quadrupole coupling constant, the asymmetry parameter, and the EFG principal axes frame (PAF) \cite{zeldes1957}. This conventional technique is named ``Zeeman perturbed NQR.'' The technique demonstrated here does not involve a static magnetic field, rather the response to the excitation direction yields the additional information, thus making the experimental setup simpler. The results are also more geometrically intuitive, the signal response is weakest along the EFG-PAF $Z-$axis, becoming zero when $\eta=0$, and largest when excitation is along EFG-PAF $X-$axis, becoming a maximum value when $\eta=1$.\\


Using such geometrical responsiveness we obtain the EFG-PAF using NQR spectroscopy of $^{35}$Cl in KClO$_3$. The Rabi frequency of such spin $3/2$ nuclei is proportional to the product of the RF magnetic field and the pulse length but also has a coefficient $\lambda$ \cite{goldman1989}, termed Rabi coefficient hereafter, that depends on the orientation of the EFG-PAF with respect to the RF excitation as well as the asymmetry in the EFG. Geometrical dependence of the Rabi frequency is exploited to determine the orientation of the EFG-PAF on a single crystal sample. The experiment utilizes a two-coil setup with a rotational nanopositioner \cite{attocube} to make goniometry measurements. A powder sample is used to compare the signal from the single crystal sample and eliminate the need to measure the field strength of the RF pulses. The experiments were performed in a cryogen-free cryostat \cite{cryostat} at temperatures ranging from $200 ~$K down to $17 ~$K. While the probe was capable of going to lower temperatures, the $T_1$ relaxation time became prohibitively large. \\

The Rabi coefficient ($\lambda$) characterizes the free induction signal generated by spin $3/2$ nuclei in a single crystal, when the same coil is used for excitation and detection as \cite{nixon2022}:
 \begin{equation}
 S \propto \lambda \sin(\lambda \Theta),
 \label{eqn:nqr signal}
 \end{equation}
where $\Theta=\frac{\sqrt{3}}{2}\gamma B_1 t_1$,   $\gamma$ is the gyromagnetic ratio, $B_1$ is the RF field strength, and $t_1$ is the excitation pulse duration \cite{nixon2022}. While the dependence of $\lambda$ on excitation direction and EFG asymmetry is somewhat involved, as detailed later, the expression for materials with almost symmetric EFG is particularly simple:
\begin{equation}
    \lambda^2 = 1 - r^2,
    \label{eqn: eta=zero}
\end{equation}
where, $r = \hat{n}\cdot\hat{Z}_{EFG}$, $\hat{n}$ is the direction of the RF excitation pulse, $\hat{Z}_{EFG}$ is the direction of the EFG-PAF $Z-$axis. Equation \ref{eqn:nqr signal} gives the condition for obtaining the maximum response $S$ for a single crystal sample. For a powder sample, the condition for maximum signal is insensitive to geometry within $1.1 \%$ \cite{xia1996}. The conditions for the maximum NQR response are compared for the powder and single crystal samples of KClO$_3$ and the Rabi coefficient is determined. EFG-PAF can subsequently be obtained through the Rabi coefficient.\\

\section{Experimental design}

\begin{figure}[h]
    \centering
    \includegraphics[width=\linewidth]{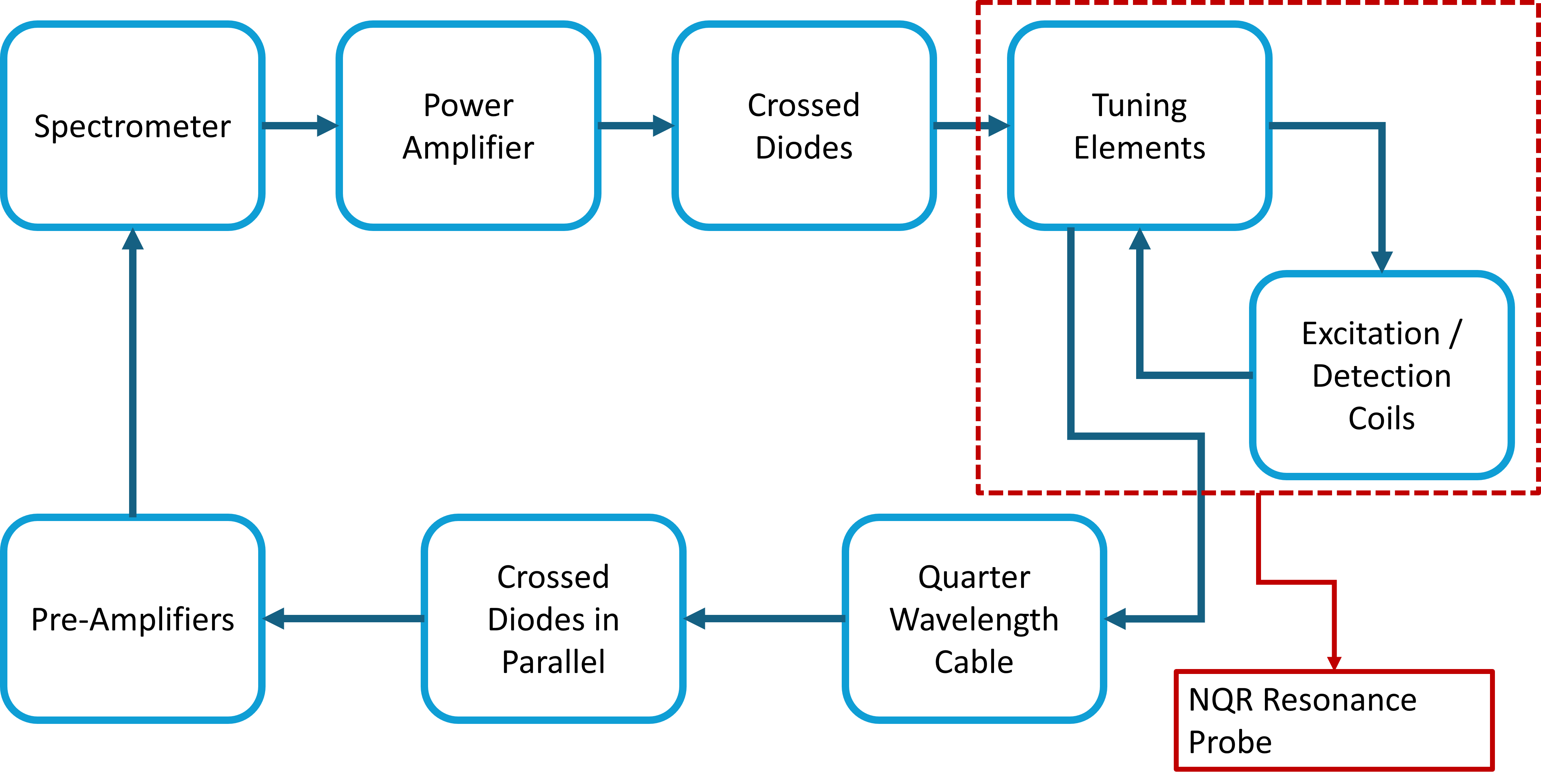}
    \caption{Block diagram of the experimental setup. The sample sits in the coils which sit inside the cryostat. Other elements are external to the cryostat.}
    \label{fig:block diagram}
\end{figure}

\begin{figure}[h]
    \centering
    \includegraphics[width=\linewidth]{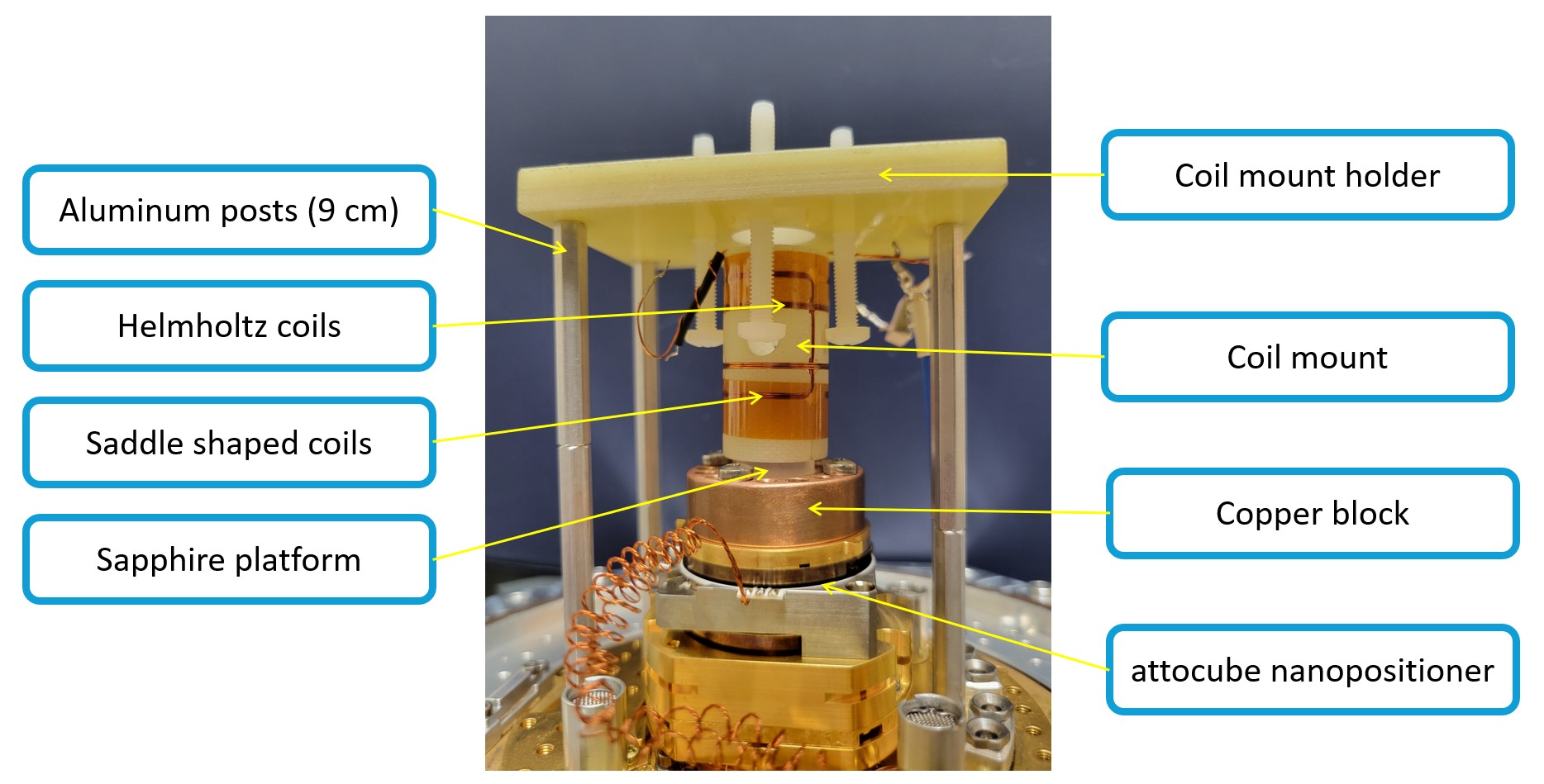}
    \caption{The NQR probe within the open sample chamber of the cryostat \cite{cryostat}. The sample sits inside a crossed pair of coils and on top of a sapphire cylinder for a strong thermal connection to the copper block mounted on top of the attocube nanopositioner. The coils are suspended with the coil mount holder using screws, to fix their position while the interior is rotated.}
    \label{fig:sample chamber}
\end{figure}

The block diagram in Fig. \ref{fig:block diagram} displays the NQR experimental setup. Pulsed excitation is generated by an RF spectrometer from Tecmag \cite{tecmag}, and amplified; after the amplification, the crossed diodes block the low-voltage noise after pulse termination. The excitation signal is then passed on to the NQR resonance probe consisting of fixed and variable capacitors as tuning elements, and the Helmholtz or saddle-shaped coils. Only one coil is used at a time, and it is used for both excitation and detection. The tuning and matching elements tune the circuit to the resonance frequency and are placed outside the cryostat sample chamber to allow re-tuning of the circuit to a changing resonance frequency at different temperatures, without disturbing the sample chamber. The two coils sit inside the sample chamber,  as shown in Fig. \ref{fig:sample chamber}. The detected NQR signal is passed on to the quarter wavelength cable. The crossed diodes turn on during the high voltage excitation pulses, shorting the quarter wavelength transmission, and preventing the damage to the pre-amplifiers. At low voltages, the diodes turn off, and pre-amplifiers amplify the signal before it is received by the spectrometer.\\

Figure \ref{fig:sample chamber} shows the setup within the sample chamber of the cryostat. When under operation, the sample chamber is evacuated and the sample is cooled through a cold finger which in turn is cooled by recirculating helium. The coils are mounted on a G-10 cylinder and suspended above the cold finger. Inside the coil mount, a sapphire platform is placed, upon which the sample sits. Since the coils are suspended by the screws, the sapphire platform is free to rotate, and the rotation is performed by an attocube nanopositioner.  A ``sniffer" coil, was placed inside the sample chamber, for diagnostic purposes. The sniffer coil detected the drop in the strength of the excitation signal due to arcing. In a conventional cryogenic cooling, helium gas flows over the sample and circuit, and arcing occurs through ionization of the helium. Somewhat surprisingly, we found that the circuit breakdown was due to vacuum arcing, in which electrons are emitted from a metallic surface \cite{slade2013}. The issue was resolved by adding polyimide tape across these metallic contacts on the attocube nanopositioner.\\ 

\begin{figure}[h]
    \centering
    \includegraphics[width=0.9\linewidth]{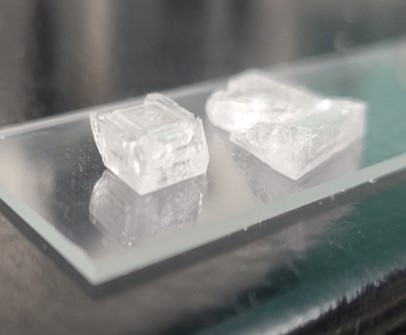}
    \caption{Homegrown potassium chlorate (KClO$_3$) single crystals.}
    \label{fig:single crystal growth}
\end{figure}

Single crystals of KClO$_3$ were grown from a super-saturated solution, heated to $45~\degree$C, adding a seed crystal, and then cooled to $20 ~\degree$C at an extremely slow rate of $1.5~\degree$C per day. Also, the evaporation was slowed down by covering the solution beaker with a perforated aluminum foil. Crystals grown with this method were of considerable size, as shown in Fig. \ref{fig:single crystal growth}. The dimensions of the crystal used were roughly $8 \times 8 \times 3~$mm$^3$ and weighed $0.36~$g. The KClO$_3$ single crystals were somewhat challenging to grow, but Ref. \cite{holden_crystals} gives a basic idea for crystal growth, while Ref. \cite{teles2015} gives a comprehensive process for growing these crystals. The powder sample was prepared in a cylindro-conical shaped mono-crystalline sapphire crucible, with an internal diameter of  $10~$mm, and maximum depth was $10~$mm, with $5 ~$mm being the depth of the cylindrical part. The powder sample had to be potted in wax to facilitate better temperature uniformity. The sample thus prepared contained $0.74 ~$g of KClO$_3$ in a $0.87 ~$g mixture of the sample and wax.  

\section{Theory}


From Eq. \ref{eqn:nqr signal}, it is evident that, for the single crystal sample, the NQR signal strength will be maximum when $\lambda \Theta = \pi/2$. For a polycrystalline or powder sample, the asymmetry parameter does not significantly affect the signal from the spin $3/2$ nuclei \cite{xia1996}. For the asymmetry parameter ($\eta$) varying from 0 to 1, the optimal flip angle $\Theta$ varies from $1.78$ to $1.80 ~$radians \cite{xia1996}. Therefore taking $\Theta = 1.79$ for maximum NQR signal for the powder sample gives less than $1 \%$ error. It is important to note that Ref. \cite{xia1996} has considered free induction decay (FID) experiments to obtain this value. However, for our powder sample, FID experiments were difficult given their shorter $T_2^*$ relaxation times. Thus, spin echo (SE) experiments were conducted for the powder sample. This will introduce a small systematic error in our calculation of $\lambda$, which can be mitigated by scaling the Rabi coefficient as shown in the following section.\\

Taking the ratio of the conditions for the maximum NQR signal strength for the single crystal and a powder sample in the same excitation field $B_1$, we get a relation between the optimal excitation pulse durations for the single crystal and powder samples and the Rabi coefficient $\lambda$, as follows:\\
\begin{equation}
    \lambda = \frac{1}{1.14} \frac{t_{pwd}}{t_{SC}},
    \label{eqn: ratio_lambda_fid}
\end{equation}
where, $t_{pwd}$ is the optimal pulse length for the powder sample and $t_{SC}$ is the optimal pulse length for the single crystal sample.

\subsection{Rabi coefficient}

\begin{figure*}[htbp]
    \centering
    \includegraphics[width=\linewidth, angle = 0]{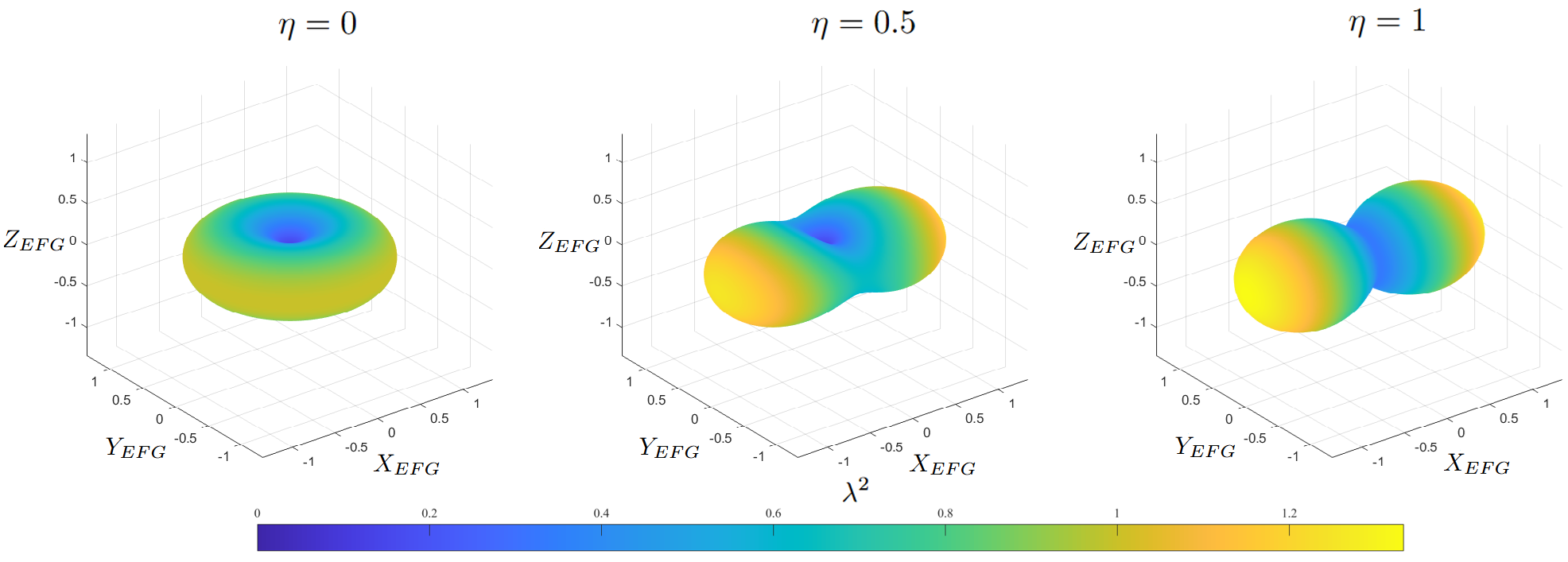}
    \caption{ Visualization of Rabi coefficient squared $\lambda^2$ for different values of the asymmetry parameter. For every RF direction with respect to the EFG-PAF, the value of $\lambda^2$ is given as the distance from the origin, creating a three-dimensional surface.} 
    \label{fig: Rabi-asymmetry}
\end{figure*}

As Ref. \cite{goldman1989} discussed, the Rabi coefficient depends on the orientation of the excitation pulse relative to the principal axes frame of the electric field gradient. A visual of the same \cite{odin1999} is shown in Fig. \ref{fig: Rabi-asymmetry}. After a few modifications, the Rabi coefficient as presented by Ref. \cite{goldman1989} can be defined by:
\begin{equation}
    \lambda^2 = \frac{2}{3} \sum_i \eta_i^2 \cdot (\hat{n}\cdot \hat{X}_i)^2,
    \label{eqn: lambda goldman}
\end{equation}
where, $\eta_i = \eta_x, \eta_y, \textrm{ and } \eta_z$, $\eta_x^2 = \frac{(V_{yy} - V_{zz}) ^2}{V_{xx}^2 + V_{yy}^2 + V_{zz}^2}$, $\eta_y^2 = \frac{(V_{zz} - V_{xx}) ^2}{V_{xx}^2 + V_{yy}^2 + V_{zz}^2}$, and $\eta_z^2 = \frac{(V_{xx} - V_{yy}) ^2}{V_{xx}^2 + V_{yy}^2 + V_{zz}^2}$, are the asymmetry parameters with respect to the EFG-PAF axes, $\hat{n}$ is the direction of the applied excitation pulse, $\hat{X}_i = \hat{X}_{EFG},$ $\hat{Y}_{EFG}$, and $\hat{Z}_{EFG}$, i.e., the respective EFG-PAF axis.\\

Using the fact that $\hat{n}$ is a unit vector, it follows from Eq. \ref{eqn: lambda goldman}:
\begin{equation}
\sum_{i} (\hat{n}\cdot\hat{X}_i)^2 = 1 ,
\end{equation}
and noting,
\begin{equation}
\sum_{i} \eta_i^2 = 3,
\end{equation}
when the RF excitation field is applied in three orthogonal directions, the sum of the squares of the Rabi coefficients must be two,
\begin{equation}
    \lambda^2_{\hat{n} = \hat{x}} + \lambda^2_{\hat{n} = \hat{y}} + \lambda^2_{\hat{n} = \hat{z}} = 2.
\label{eqn: lambda_squared_sum}
\end{equation}
This conservation law turns out to be useful for correction of any systematic errors.\\

Although elegant, Eq. \ref{eqn: lambda goldman} contains certain redundant terms, eliminating those and retaining the independent terms,
\begin{equation}
    \lambda^2 = \frac{3}{3 + \eta^2}[(1 - r^2)(1 + \frac{2}{3}\eta - \frac{1}{3}\eta^2) - \frac{4}{3}q^2 \eta + \frac{4}{9}\eta^2],
    \label{eqn: lambda ind. comp.}
\end{equation}
where, $q = \hat{n}\cdot\hat{Y}_{EFG}$ is a directional cosine. Looking at the two extrema of Eq. \ref{eqn: lambda ind. comp.}, i.e., $\eta  = 1$, and $\eta = 0$, we get the following results. For $\eta = 1$, the case of maximum asymmetry:
\[ \lambda^2 = p^2 + \frac{1}{3},\]
where, $p^2 = (\hat{n}\cdot \hat{X}_{EFG})^2 = 1 - q^2 - r^2$, the Rabi coefficient in this case will be maximum when the excitation direction $\hat{n}$ is parallel to the $\hat{X}_{EFG}$ direction, and minimum when they are perpendicular to each other. For $\eta = 0$, the case of no asymmetry, Eq. \ref{eqn: eta=zero} follows. The Rabi coefficient in this case will be maximum when the excitation direction $\hat{n}$ is perpendicular to the $\hat{Z}_{EFG}$ direction and minimum when they are parallel to each other. The case of $\eta = 0$ is especially important as a lot of materials have very low asymmetry \cite{choudhary2020}.

\subsection{Determining EFG-PAF through the Rabi coefficient}

\begin{figure}[h]
    \centering
    \includegraphics[width=\linewidth]{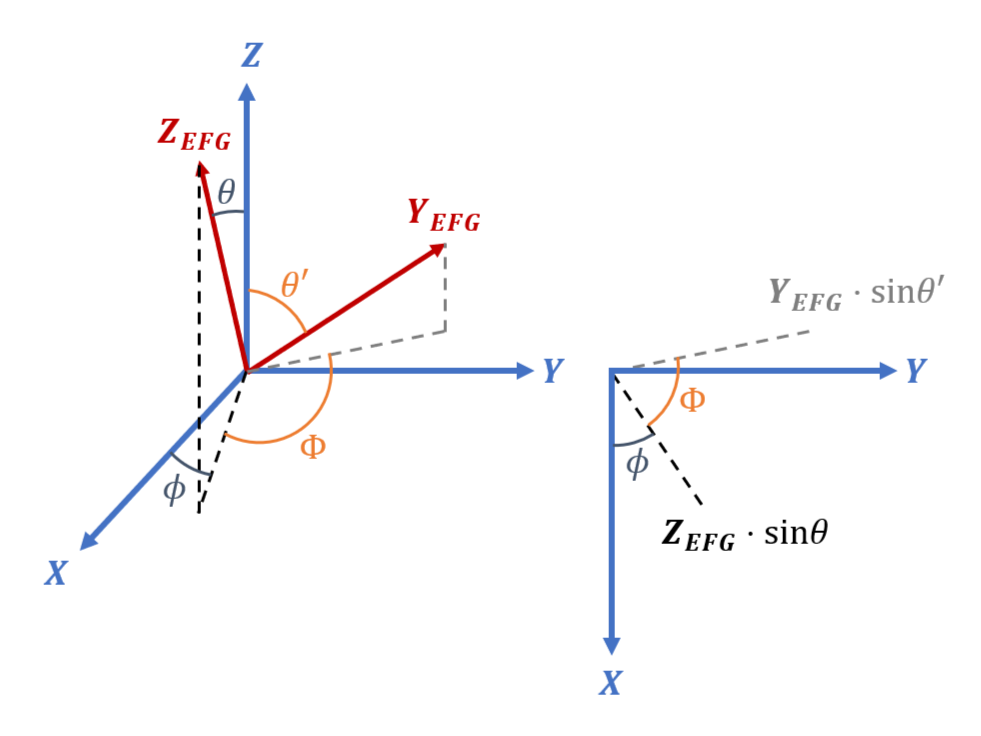}
    \caption{The orientation of EFG-PAF with the lab frame. The figure on the right shows the top view.}
    \label{fig:EFG_axes}
\end{figure}

Using Eq. \ref{eqn: ratio_lambda_fid} and Eq. \ref{eqn: lambda ind. comp.}, we can determine the Rabi coefficient $\lambda$ for our system, and through that, the orientation of EFG-PAF can be obtained. Consider the orientation of EFG-PAF with respect to the lab axes frame as given in Fig. \ref{fig:EFG_axes}. The Helmholtz coils and the saddle coils shown in Fig. \ref{fig:sample chamber}, generate the excitation pulse signal in the lab frame $\hat{Z}$ and $\hat{X}$ directions respectively. The saddle coil can be initially aligned with one of the crystal axes, in our case the axis $a$ of KClO$_3$. The attocube positioner \cite{attocube} rotates the sample, the axis of rotation being the ${Z}-$axis, thus rotating the sample relative to the saddle coil field. \\

As shown in Fig. \ref{fig:EFG_axes}, the $Z_{EFG}-$axis makes an angle of $\theta$ with the lab frame $Z-$axis, and an angle of $\phi$ with the lab frame $X-$axis in the $XY-$plane, we can say:
\begin{equation}
    \hat{Z}_{EFG}= \sin\theta \cos\phi\hat{X} + \sin\theta \sin\phi\hat{Y} + \cos\theta\hat{Z}
    \label{eqn: EFG_Z}  . 
\end{equation}
Similarly, for the ${Y}_{EFG}-$axis, let $\theta'$ be the angle that it makes with the $Z-$axis, and let $\Phi + \phi$ be the angle the projection of ${Y}_{EFG}-$axis in the $XY$ plane makes with the $X-$axis, we can say:

\begin{equation}
    \hat{Y}_{EFG}= \sin\theta' \cos(\phi + \Phi)\hat{X} + \sin\theta' \sin(\phi + \Phi)\hat{Y} + \cos\theta'\hat{Z}.
    \label{eqn: EFG_X}
\end{equation}\\
Since the PAF is an orthogonal system, $\hat{Y}_{EFG} \cdot \hat{Z}_{EFG} = 0$, it gives a constraint on these angles as, $\cos\Phi=-\cot\theta'\cdot\cot\theta.$\\

For Helmholtz coils, $\hat{n} = \hat{Z},$\\\\
$q^2 = (\hat{Z} \cdot \hat{Y}_{EFG})^2 = \cos^2 \theta',$ and\\\\
$r^2 = (\hat{Z} \cdot \hat{Z}_{EFG})^2 = \cos^2\theta.$ Therefore,
\begin{equation}
    \begin{split}
        \lambda_H^2 = \frac{3}{3 + \eta^2}[\frac{4}{9}\eta^2 + (1 - \cos ^2\theta)(1 + \frac{2}{3}\eta - \frac{1}{3}\eta^2) \\ - \frac{4}{3}\cos^2 \theta' \eta].
    \end{split}
    \label{eqn: lambda_helm_long}
\end{equation}
For saddle coils, $\hat{n} = \hat{X}$,\\\\
$q^2 = (\hat{X} \cdot \hat{Y}_{EFG})^2 = \sin^2 \theta' \cos^2 (\phi+\Phi),$ and\\\\
$r^2 = (\hat{X} \cdot \hat{Z}_{EFG})^2 = \sin ^2\theta \cos^2 \phi.$
Therefore,
\begin{equation}
    \begin{split}
        \lambda_S^2 = \frac{3}{3 + \eta^2}[\frac{4}{9}\eta^2 + (1 - \sin ^2\theta \cos^2 \phi)(1 + \frac{2}{3}\eta - \frac{1}{3}\eta^2)\\ - \frac{4}{3}\sin^2 \theta' \cos^2 (\phi+\Phi) \eta]    ,
    \end{split}
    \label{eqn: lambda_saddle_long}    
\end{equation}
where $\lambda_H$, and $\lambda_S$ are the Rabi coefficient determined for the Helmholtz and saddle coils respectively, angle $\phi$ is a varied quantity, different values of $\phi$ can be assigned to different positions of a single crystal with respect to the lab frame ${X}-$axis. It can be shown that $\lambda_S^2$ can be written as a constant and a double angle sinusoidal function of $\phi$, with independent variables $\eta,$ $\theta,$ and $\Phi$. A second orthogonal rotation would be required to determine all three variables.
\\

For the special case of $\eta = 0$, from Eq. \ref{eqn: lambda_helm_long} and Eq. \ref{eqn: lambda_saddle_long}:
\begin{equation}
 \lambda_H^2  =   \sin^2\theta,   
\label{eqn: Rabi_Helm}
\end{equation}
\begin{equation}
\lambda_S^2 = 1 -  \sin ^2\theta \cos^2 \phi.    
\label{eqn: Rabi_Saddle}
\end{equation}
From Eq. \ref{eqn: Rabi_Saddle}, we note that for $\phi = 0$ or $\pi$, there will be a minimum value of $\lambda_S^2$ ($\lambda^2_S|_{min}$). The crystal's orientation in the lab frame, corresponding to these conditions, gives the angle between the relevant axes, in our case $a$ and the projection of the ${Z}_{EFG}-$axis on to the $XY$ plane. In addition, the value at these locations, $\lambda^2_S|_{min}$ can be used to determine the angle $\theta$, to fully define $\hat{Z}_{EFG}$ with respect to the single crystal. Furthermore,
\begin{equation}
    \lambda_H ^2 + \lambda_{S}^2|_{min} = \lambda_{S}^2 |_{max} = 1,
\label{eqn: zero eta rabi condition}
\end{equation}
where, $  \lambda_{S}^2 |_{max}$ is the maximum value of $\lambda_S^2$. If Eq. \ref{eqn: zero eta rabi condition} is satisfied, it is indicative that the material has zero asymmetry. It is clear that, for very low asymmetry, only one rotation is sufficient to determine the EFG-PAF.

\section{Results and Discussion}

\subsection{Finding the EFG-$Z$ Axis}


For KClO$_3$ the asymmetry parameter $\eta$ is close to zero \cite{zeldes1957}. Therefore, it is possible to find the orientation by using just the saddle coils; however, the Helmholtz coils are also used, as a confirmational test. 
Then, by understanding the relationship between the growth of the single crystal and the crystallographic axes, the EFG-PAF orientation can be given with respect to the crystal axes as well.\\



Rabi coefficient was determined using the optimal pulse length observed for the maximum signal output from FID and SE experiments for the single crystal and powder sample respectively. The powder sample required SE experiments due to the very short $T_2^*$ relaxation times, and experimental dead-time due to ringing. For SE experiments, only the excitation pulse was varied, whereas the refocusing pulse remained the same for all the data points. Phase cycling was used to suppress ringing artifacts and isolate the echo.\\

\begin{figure}[h]
\includegraphics[width=0.75\linewidth]{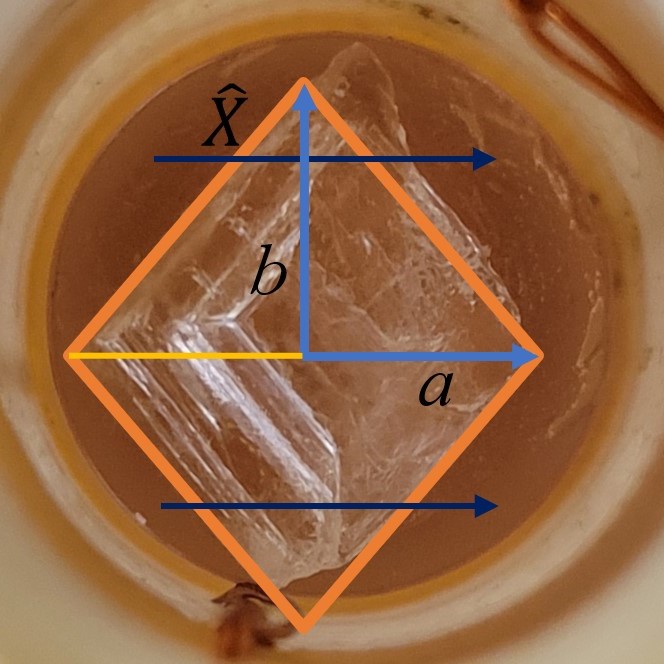}
\caption{The yellow line shows the projection of ${Z}_{EFG}-$axis onto the single crystal. The dark blue lines denote the saddle-shaped coil field direction $\hat{X}$. The orange diamond shows the flake in `$ab$' crystallographic plane, `$a$' and `$b$' axes are determined using Ref. \cite{groth_german}. The sample rests on a cylindrical sapphire platform of diameter 14.1 mm.}
\label{fig:Z-EFG_location}
\end{figure}

\begin{figure}[h]
\includegraphics[width=\linewidth]{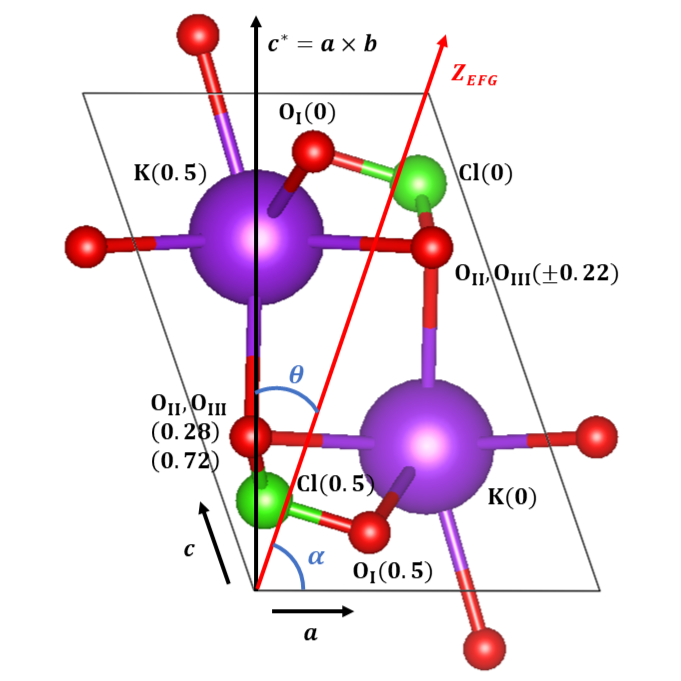}
\caption {The ${Z}_{EFG}-$axis visualized  on the atomic positions of KClO$_3$ structure projected onto the $b$-face. Angles $\theta$ and $\alpha$ are the angles ${Z}_{EFG}-$axis makes with crystallographic $c^*$ axis and $a$ axis respectively. The $c^*-$axis is obtained by performing a cross product of $a$ and $b$, and is parallel to the lab frame ${Z}-$axis. Numbers in parenthesis are distances of atoms above and below the $ac$ plane in fractional $b-$axis units obtained from Ref. \cite{schlick1972}.}
\label{fig: kclo3_vesta}
\end{figure}

Performing the rotation of the sample and finding the minimum of the Rabi coefficient squared for the saddle coils ($\lambda_S^2$), we find the position of the single crystal which corresponds to $\phi = 0\degree.$ Figure \ref{fig:Z-EFG_location} shows the single crystal in that orientation. By understanding the relationship of crystal growth with respect to the crystallographic axes \cite{groth_german}, the crystal axes and the $ab$ crystallographic plane are marked. Since it is now known that the ${Z}_{EFG}-$axis makes an angle of $0\degree$ or $180\degree$ with the $\hat{X}$ field direction when the crystal is in the position shown in Fig. \ref{fig:Z-EFG_location}, the yellow line shows the projection of ${Z}_{EFG}-$axis on the face of the crystal. It is also known that the ${Z}_{EFG}-$axis makes the angle $\theta$ with the $c^*$ crystal axis $-$ obtained by performing a cross product $a \times b,$ and lies in the $ac$ crystallographic plane, as shown in Fig. \ref{fig: kclo3_vesta}. The crystal axis $c^*$ is parallel to the laboratory frame direction $\hat{Z}$. Combining the results, the projection of the ${Z}_{EFG}-$axis is determined to be along the yellow line.\\

\begin{figure}[h]
\includegraphics[width=\linewidth]{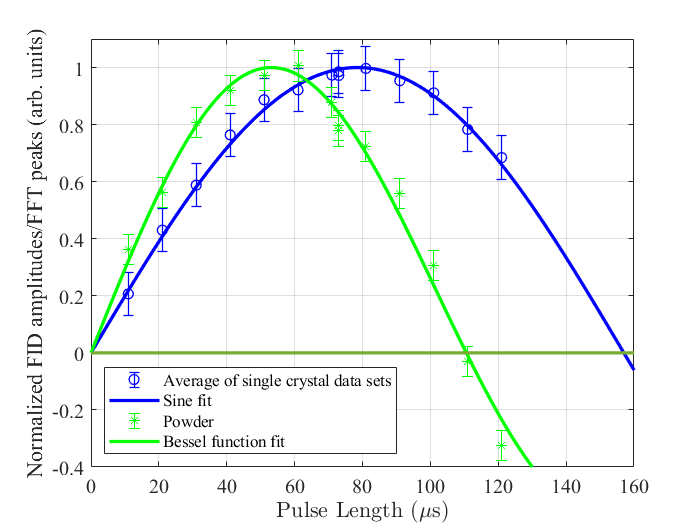}
\caption{Single crystal FID and powder SE plots for Helmholtz coils fit to sine curve and $J_1$ Bessel function \cite{nixon2022} respectively. Reduced $\chi^2 = 0.1$, and $0.9$ for the single crystal sample and the powder sample respectively. Experiments conducted at $ 200~$K.}
\label{fig:helm_compare}
\end{figure}

\begin{figure}[h]
\includegraphics[width=\linewidth]{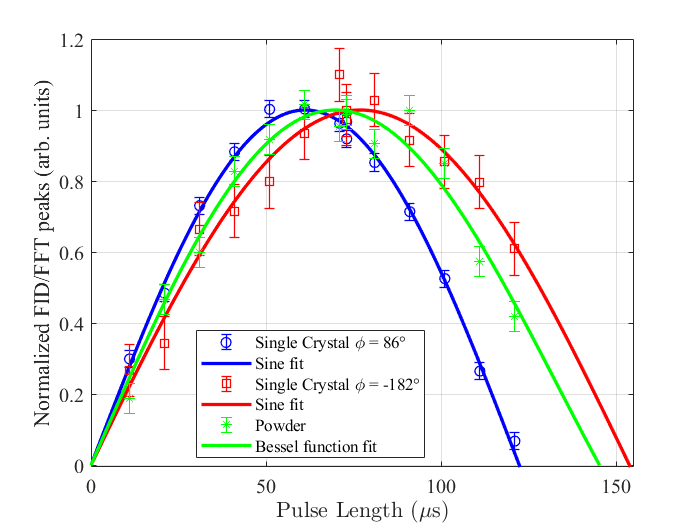}
\caption{Single crystal FID and powder SE plots for saddle coils fit to sine curve and $J_1$ Bessel function \cite{nixon2022} respectively. Reduced $\chi^2 = 0.9, 0.5,$ and $1.6$ for $\phi = 86\degree, \phi = -182\degree,$ and powder sample respectively. Experiments conducted at $200~$K.}
\label{fig:saddle_compare}
\end{figure}

The plots comparing the experiments for both the samples are shown in Fig. \ref{fig:helm_compare} and Fig. \ref{fig:saddle_compare} for the excitation field generated by the Helmholtz coils and the saddle coils respectively. Figure \ref{fig:helm_compare} contains the plot of powder sample's SE data and averaged single crystal data sets, since no significant variation in the signal is observed with $\phi$ when using the Helmholtz coils. Figure \ref{fig:saddle_compare} contains the plot for two orientations of the single crystal, which have the largest and the smallest observed values for the optimal pulse lengths, along with powder sample's SE data. It is evident from Fig. \ref{fig:saddle_compare} that the optimal pulse lengths vary significantly, reinforcing the geometric dependence of the Rabi coefficient.\\

\begin{figure}[h]
\includegraphics[width=\linewidth]{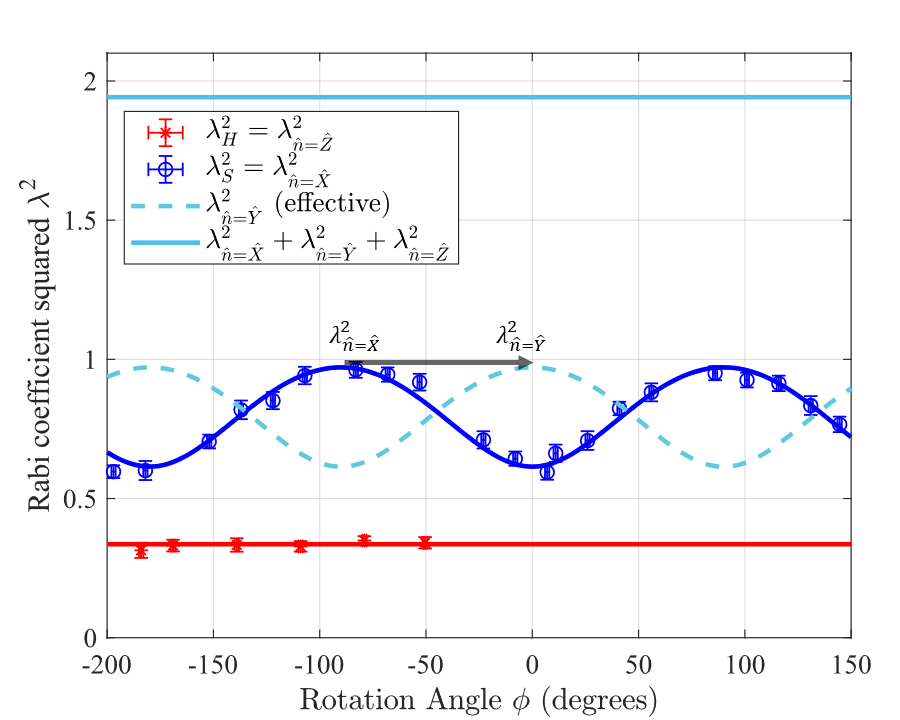} 
\caption{Rabi coefficient squared with the $\phi$ rotation angles for saddle-shaped ($\lambda_S^2 = \lambda_{\hat{n}=\hat{X}}^2$) and Helmholtz coils ($\lambda_H^2 = \lambda_{\hat{n}=\hat{Z}}^2$), the $X$ error bars are within the symbol. Reduced $\chi^2 = 0.8$ and $0.4$ respectively. Effective $\lambda_{\hat{n}=\hat{Y}}^2$ is plotted by shifting  $\lambda_{\hat{n}=\hat{X}}^2$ by $90\degree$. The Helmholtz coil data is fit to Eq. \ref{eqn: Rabi_Helm} and the saddle coil data is fit to Eq. \ref{eqn: Rabi_Saddle}. The sum of the fit of orthogonal components is a constant as discussed in the text. Experiments conducted at $200~$K.}
\label{fig:rabi-rotation}
\end{figure}


We see from Eq. \ref{eqn: Rabi_Helm} that the Rabi coefficient does not vary with the angle $\phi$ for the Helmholtz coils, and from Eq. \ref{eqn: Rabi_Saddle}, we see that the Rabi coefficient does vary with the angle $\phi$ for the saddle-shaped coils. Figure \ref{fig:rabi-rotation} confirms the behavior predicted in Eq. \ref{eqn: Rabi_Helm} and Eq. \ref{eqn: Rabi_Saddle}. We use the optimal pulse lengths of the powder and single crystal samples for the Helmholtz coils and the saddle coils, and find the Rabi coefficient using Eq. \ref{eqn: ratio_lambda_fid}. From the data fits in the Rabi coefficient plots in Fig. \ref{fig:rabi-rotation}, we get:    
\begin{equation}
    \lambda_H ^2 + \lambda_{S} |_{min}^2 = 0.96 \pm 0.02,
    \label{eqn: lambda_H + lambda_S_min}
\end{equation}
\begin{equation}
    \lambda_{S} |_{max}^2 = 0.98 \pm 0.01,
    \label{eqn: lambda_S_max}
\end{equation}
\begin{equation}
    \lambda^2_{\hat{n} = \hat{x}} + \lambda^2_{\hat{n} = \hat{y}} + \lambda^2_{\hat{n} = \hat{z}} = 1.94 \pm 0.03,
    \label{eqn: lambda_sq sum experimental}
\end{equation}
thus, Eq. \ref{eqn: zero eta rabi condition} and Eq. \ref{eqn: lambda_squared_sum} are reasonably satisfied. Although, notably all three values are smaller than expected, suggesting a small but systematic error to the experiment. There could be multiple small systematic errors. One possible cause of systematic error could be the shape and size of the powder sample versus the single crystal within the coils experiencing slightly different RF field strengths. Another systematic error could arise from the use of SE pulses for the powder sample calibration. The optimal excitation $\Theta$ for a powder sample excited with a $\Theta-2\Theta$ SE sequence is slightly smaller than for an FID \cite{nixon2022} and would therefore tend to underestimate $\lambda$ when using Eq. \ref{eqn: ratio_lambda_fid}.\\


The systematic errors can be mitigated by scaling the summation of $\lambda^2$ value to two, as noted in Eq. \ref{eqn: lambda_squared_sum}. All other values are scaled by the same amount. Fitting the Helmholtz coils data to Eq. \ref{eqn: Rabi_Helm}, we get,
\[\theta = 36.0^{\circ} \pm 0.5 \degree,\]
and, from fits of the saddle coil data to Eq. \ref{eqn: Rabi_Saddle} we get,
\[\theta = 36.9\degree \pm 1.0\degree,\]
Considering the weighted mean of the above results for $\theta$, we get,
\begin{equation}
\theta = 36.2\degree \pm 0.5\degree.    
\end{equation}
This angle gives the position of the ${Z}_{EFG}-$axis with respect to the single crystal sample as well. 
\\

The angle between ${Z}_{EFG}-$axis and the `$ab$' crystallographic plane is thus $\alpha = 53.8\degree \pm 0.5\degree$, in reasonable agreement with previous works. Reference \cite{zeldes1957} obtains $\alpha = 52.1 \degree \pm 0.5 \degree$ at room temperature. It is hypothesized that the ${Z}_{EFG}-$axis should be normal to the O$_3$ plane which estimates $\alpha $ to be between $51.9 \degree$ \cite{danielsen1981} and $55.7\degree$ \cite{zeldes1957}.\\

\begin{figure}[h]
    \centering
    \includegraphics[width=\linewidth]{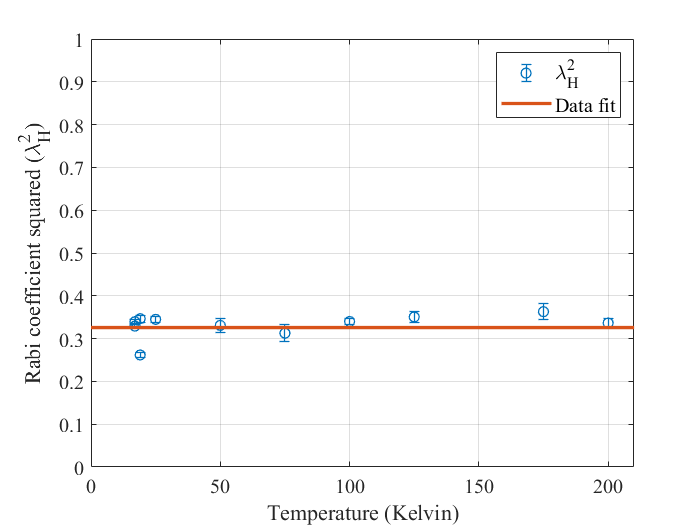}
    \caption{Rabi coefficient squared at different temperatures for the Helmholtz coils.}
    \label{fig:lambda_H vs Temp}
\end{figure}

The Rabi coefficient was measured for temperatures down to $17$ K, using the Helmholtz coils. The temperature of the sample was determined by the resonance frequency \cite{utton1967}. No significant variation in the Rabi coefficient was observed. This implies that the orientation of the principal axes frame of the EFG was largely insensitive to the temperature for KClO$_3$. However, as is seen in Fig. \ref{fig:lambda_H vs Temp} there is an outlier at $T = 19$ K. As the temperature decreases, the quality factor of the circuit increases, at $17~$K, the quality factor was $51$, almost $1.6$ times of the value of $32$ at $200 ~$K. This drastic increase in quality factor narrows the tuning frequency range for resonance and makes it difficult to tune the circuit at lower temperatures. 

\subsection{The relaxation times}

\begin{figure}[h]
    \centering
    \includegraphics[width=\linewidth]{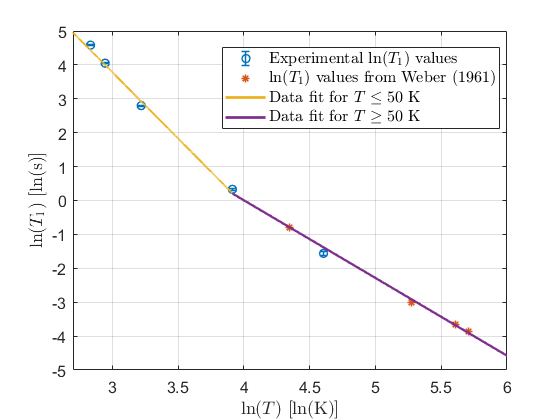}
    \caption{Experimental and literature \cite{weber1961} values for spin-lattice relaxation time $T_1$ plotted with the temperature. The data is fit to $\ln(T_1) = a\ln(T) + b$. For $T \ge 50$ K, $a = -2.3$. For $T\le 50$ K, $ a = -3.9$.}
    \label{fig:t1_vs_temp}
\end{figure}

The relaxation times $T_1$ and $T_2^*$ for the sample were measured at multiple different temperatures, ranging from $17$ K to $200$ K. The $T_1$ relaxation time increases drastically as the temperature decreases. The behavior is displayed in Fig. \ref{fig:t1_vs_temp}, and is consistent with the trends observed previously for a temperature range of $77 - 300~$K \cite{weber1961}. The logarithm of the relaxation time is plotted against the logarithm of the temperature to determine the power dependence of $T_1$ on the temperature. As seen in Fig. \ref{fig:t1_vs_temp}, for temperatures above $50~$K, the data fits to a power of $-2.3$, and for temperatures below $50~$K, the data fits to a power of $-3.9$.\\

Multiple theories have been developed to describe the temperature dependence of the spin-lattice relaxation for molecular crystals with quadrupole coupling. Reference \cite{bayer1951} and Ref. \cite{ayant1956}  proposed a molecular torsional oscillator model, and Ref. \cite{vankranendonk1954} proposed a model in which lattice vibrations are responsible. However, later developments suggest that the molecular torsional oscillator model is a better fit for the higher temperatures. For a temperature range of $77 - 350~$K, Ref. \cite{ramesh2008} proposed the molecular torsional oscillator theory and obtained $T_1 \propto T^{-2.31}$ for NaClO$_3$, which is a material whose properties closely align with KClO$_3$. In paradichlorobenzene for $^{35}$Cl nuclei, Ref. \cite{woessner1963} observed $T_1 \propto T^{-2.38}$ in the temperature range $77 - 220~$K, and at room temperature Ref. \cite{weber1961} concluded that only the molecular torsional oscillator model could account for the observed relaxation behavior.\\

However, for temperatures lower than $50~$K, it appears that the molecular torsional oscillations freeze out and do not influence the relaxation mechanism as much. For temperatures $T > 0.5\Theta_D$, where $\Theta_D$ is the Debye temperature, the relaxation time $T_1$ is predicted to have a power law of `-2', but the power transitions to `-7' for $T < 0.02\Theta_D$ \cite{vankranendonk1954}. The Debye temperature of KClO$_3$ is estimated to be $\Theta_D = 232~$K \cite{korabelnikov2016}, thus $0.02\Theta_D$ would be at $4.64 ~$K, which is much less than the lowest temperature data measured here at $17 ~$K. However, the data fitting to $T_1 \propto T^{-3.9}$ at $T \le 50 ~$K suggests that at low temperatures the molecular torsional oscillations freeze out and the lattice vibrations dominate the relaxations.\\

The relaxation time $T_2^*$ was observed for temperatures ranging from room temperature to $17 ~$K. Although there were some variations in the observed relaxation time $T_2^*$, the values remained close to $0.7 ~$ms at either extrema of the temperature range studied, consistent with the value observed by Ref. \cite{weber-hahn1960} for the single crystal sample. For the powder sample, $T_2^*$ was observed to be $0.2 ~$ms, over the same range, with minor variations. Reference \cite{melnick1982} observed an increase in $T_2^*$ with decrease in the temperature for KClO$_3$ powder sample, from about $0.25 ~$ms at room temperature to about $0.4 ~$ms at $77 ~$K. This discrepency can be explained by the fact that $T_2^*$ is often dominated by the inhomogeneity in the electric field gradient which can vary depending on how the powder was manufactured \cite{malone2011}. Thus, it is unsurprising that the $T_2^*$ reported by Ref. \cite{melnick1982} is different from what was observed here.

\section{Conclusion}

We have demonstrated a straightforward methodology to determine the EFG-PAF for spin $3/2$ nuclei in crystalline material, using the Rabi coefficient $\lambda$. The geometric dependency of the Rabi coefficient with the excitation direction and the asymmetry parameter ($\eta$) can be exploited and the relation between the optimal flip angles and the optimal excitation pulse duration gives a simple relationship, especially for materials which have a negligible asymmetry parameter $\eta$. In general, it is a non-trivial task to calculate the magnetic field strength of the excitation coils accurately. We use, however, the NQR response from a powder sample to calibrate the field for the single crystal. In addition, symmetry rules with respect to $\lambda^2$ can also serve as a calibration check. \\

It was found that there are two competing theories explaining the spin-lattice relaxation mechanism in molecular crystals with quadrupole coupling and we find that both of them are applicable in KClO$_3$, but at different temperature ranges. For our sample, the molecular torsional oscillator model explains the behavior witnessed at temperatures above $50$ K, and below $50$ K, the oscillations start to freeze out and the lattice vibrations contribute more towards the relaxation mechanism.\\

All NQR experiments were successfully performed in a cryogen-free cryostat. The operation of the cryostat presented no additional noise, with experiments done at temperatures as low as $17$ K. Some issues arose due to the strong excitation pulses, including both vacuum arcing and heating; solutions included refinement of the probe, sample and pulse sequence modifications. Given the global shortage of helium, the successful operation of NQR at cryogenic temperatures in a cryogen-free cryostat could open up new arenas of applications for this zero-field NMR technique. 
   
\section{Acknowledgments}

The authors thank Dr. Mingzhen Tian for letting us use her laboratory space, Dr. Xiaoyan Tan, and Dr. Zachary Messegee for their help growing the single crystal sample, and Sándor Nyerges for machining the various components of the NQR probe. The support for this project comes from the National Science Foundation award number 2214194.

\bibliography{main}

\begin{thebibliography}{57}%
\makeatletter
\providecommand \@ifxundefined [1]{%
 \@ifx{#1\undefined}
}%
\providecommand \@ifnum [1]{%
 \ifnum #1\expandafter \@firstoftwo
 \else \expandafter \@secondoftwo
 \fi
}%
\providecommand \@ifx [1]{%
 \ifx #1\expandafter \@firstoftwo
 \else \expandafter \@secondoftwo
 \fi
}%
\providecommand \natexlab [1]{#1}%
\providecommand \enquote  [1]{``#1''}%
\providecommand \bibnamefont  [1]{#1}%
\providecommand \bibfnamefont [1]{#1}%
\providecommand \citenamefont [1]{#1}%
\providecommand \href@noop [0]{\@secondoftwo}%
\providecommand \href [0]{\begingroup \@sanitize@url \@href}%
\providecommand \@href[1]{\@@startlink{#1}\@@href}%
\providecommand \@@href[1]{\endgroup#1\@@endlink}%
\providecommand \@sanitize@url [0]{\catcode `\\12\catcode `\$12\catcode `\&12\catcode `\#12\catcode `\^12\catcode `\_12\catcode `\%12\relax}%
\providecommand \@@startlink[1]{}%
\providecommand \@@endlink[0]{}%
\providecommand \url  [0]{\begingroup\@sanitize@url \@url }%
\providecommand \@url [1]{\endgroup\@href {#1}{\urlprefix }}%
\providecommand \urlprefix  [0]{URL }%
\providecommand \Eprint [0]{\href }%
\providecommand \doibase [0]{https://doi.org/}%
\providecommand \selectlanguage [0]{\@gobble}%
\providecommand \bibinfo  [0]{\@secondoftwo}%
\providecommand \bibfield  [0]{\@secondoftwo}%
\providecommand \translation [1]{[#1]}%
\providecommand \BibitemOpen [0]{}%
\providecommand \bibitemStop [0]{}%
\providecommand \bibitemNoStop [0]{.\EOS\space}%
\providecommand \EOS [0]{\spacefactor3000\relax}%
\providecommand \BibitemShut  [1]{\csname bibitem#1\endcsname}%
\let\auto@bib@innerbib\@empty
\bibitem [{\citenamefont {Grant}(1969)}]{grant1969}%
  \BibitemOpen
  \bibfield  {author} {\bibinfo {author} {\bibfnamefont {R.~W.}\ \bibnamefont {Grant}},\ }\bibfield  {title} {\bibinfo {title} {{Nuclear Electric Field Gradient at the Iron Sites in Ca$_2$Fe$_2$O$_5$ and Ca$_2$FeAlO$_5$}},\ }\href {https://doi.org/10.1063/1.1672117} {\bibfield  {journal} {\bibinfo  {journal} {The Journal of Chemical Physics}\ }\textbf {\bibinfo {volume} {51}},\ \bibinfo {pages} {1156} (\bibinfo {year} {1969})}\BibitemShut {NoStop}%
\bibitem [{\citenamefont {Colville}(1970)}]{colville1970}%
  \BibitemOpen
  \bibfield  {author} {\bibinfo {author} {\bibfnamefont {A.~A.}\ \bibnamefont {Colville}},\ }\bibfield  {title} {\bibinfo {title} {{The crystal structure of Ca${\sb 2}$Fe${\sb 2}$O${\sb 6}$ and its relation to the nuclear electric field gradient at the iron sites}},\ }\href {https://doi.org/10.1107/S056774087000434X} {\bibfield  {journal} {\bibinfo  {journal} {Acta Crystallographica Section B}\ }\textbf {\bibinfo {volume} {26}},\ \bibinfo {pages} {1469} (\bibinfo {year} {1970})}\BibitemShut {NoStop}%
\bibitem [{\citenamefont {Iglesias}\ \emph {et~al.}(2001)\citenamefont {Iglesias}, \citenamefont {Schwarz}, \citenamefont {Blaha},\ and\ \citenamefont {Baldomir}}]{iglesias2001}%
  \BibitemOpen
  \bibfield  {author} {\bibinfo {author} {\bibfnamefont {M.}~\bibnamefont {Iglesias}}, \bibinfo {author} {\bibfnamefont {K.}~\bibnamefont {Schwarz}}, \bibinfo {author} {\bibfnamefont {P.}~\bibnamefont {Blaha}},\ and\ \bibinfo {author} {\bibfnamefont {D.}~\bibnamefont {Baldomir}},\ }\bibfield  {title} {\bibinfo {title} {{Electronic structure and electric field gradient calculations of Al$_2$SiO$_5$ polymorphs}},\ }\href {https://doi.org/10.1007/s002690000123} {\bibfield  {journal} {\bibinfo  {journal} {Physics and Chemistry of Minerals}\ }\textbf {\bibinfo {volume} {28}},\ \bibinfo {pages} {67} (\bibinfo {year} {2001})}\BibitemShut {NoStop}%
\bibitem [{\citenamefont {Vojvodin}\ \emph {et~al.}(2022)\citenamefont {Vojvodin}, \citenamefont {Holmes}, \citenamefont {Watanabe}, \citenamefont {Rawson},\ and\ \citenamefont {Schurko}}]{vojvodin2022}%
  \BibitemOpen
  \bibfield  {author} {\bibinfo {author} {\bibfnamefont {C.~S.}\ \bibnamefont {Vojvodin}}, \bibinfo {author} {\bibfnamefont {S.~T.}\ \bibnamefont {Holmes}}, \bibinfo {author} {\bibfnamefont {L.~K.}\ \bibnamefont {Watanabe}}, \bibinfo {author} {\bibfnamefont {J.~M.}\ \bibnamefont {Rawson}},\ and\ \bibinfo {author} {\bibfnamefont {R.~W.}\ \bibnamefont {Schurko}},\ }\bibfield  {title} {\bibinfo {title} {{Multi-component crystals containing urea: mechanochemical synthesis and characterization by $^{35}$Cl solid-state NMR spectroscopy and DFT calculations}},\ }\href {https://doi.org/10.1039/d1ce01610e} {\bibfield  {journal} {\bibinfo  {journal} {CrystEngComm}\ }\textbf {\bibinfo {volume} {24}},\ \bibinfo {pages} {2626} (\bibinfo {year} {2022})}\BibitemShut {NoStop}%
\bibitem [{\citenamefont {Hartman}\ and\ \citenamefont {Capistran}(2024)}]{hartman2024}%
  \BibitemOpen
  \bibfield  {author} {\bibinfo {author} {\bibfnamefont {J.~D.}\ \bibnamefont {Hartman}}\ and\ \bibinfo {author} {\bibfnamefont {D.}~\bibnamefont {Capistran}},\ }\bibfield  {title} {\bibinfo {title} {{Predicting $^{51}$V nuclear magnetic resonance observables in molecular crystals}},\ }\href {https://doi.org/https://doi.org/10.1002/mrc.5420} {\bibfield  {journal} {\bibinfo  {journal} {Magnetic Resonance in Chemistry}\ }\textbf {\bibinfo {volume} {62}},\ \bibinfo {pages} {416} (\bibinfo {year} {2024})}\BibitemShut {NoStop}%
\bibitem [{\citenamefont {Christiansen}\ \emph {et~al.}(1976)\citenamefont {Christiansen}, \citenamefont {Heubes}, \citenamefont {Keitel}, \citenamefont {Klinger}, \citenamefont {Loeffler}, \citenamefont {Sandner},\ and\ \citenamefont {Witthuhn}}]{christiansen1976}%
  \BibitemOpen
  \bibfield  {author} {\bibinfo {author} {\bibfnamefont {J.}~\bibnamefont {Christiansen}}, \bibinfo {author} {\bibfnamefont {P.}~\bibnamefont {Heubes}}, \bibinfo {author} {\bibfnamefont {R.}~\bibnamefont {Keitel}}, \bibinfo {author} {\bibfnamefont {W.}~\bibnamefont {Klinger}}, \bibinfo {author} {\bibfnamefont {W.}~\bibnamefont {Loeffler}}, \bibinfo {author} {\bibfnamefont {W.}~\bibnamefont {Sandner}},\ and\ \bibinfo {author} {\bibfnamefont {W.}~\bibnamefont {Witthuhn}},\ }\bibfield  {title} {\bibinfo {title} {{Temperature dependence of the electric field gradient in noncubic metals}},\ }\href {https://doi.org/10.1007/BF01312998} {\bibfield  {journal} {\bibinfo  {journal} {{Zeitschrift f{\"u}r Physik B Condensed Matter}}\ }\textbf {\bibinfo {volume} {24}},\ \bibinfo {pages} {177} (\bibinfo {year} {1976})}\BibitemShut {NoStop}%
\bibitem [{\citenamefont {Nishiyama}\ \emph {et~al.}(1976)\citenamefont {Nishiyama}, \citenamefont {Dimmling}, \citenamefont {Kornrumpf},\ and\ \citenamefont {Riegel}}]{nishiyama1976}%
  \BibitemOpen
  \bibfield  {author} {\bibinfo {author} {\bibfnamefont {K.}~\bibnamefont {Nishiyama}}, \bibinfo {author} {\bibfnamefont {F.}~\bibnamefont {Dimmling}}, \bibinfo {author} {\bibfnamefont {T.}~\bibnamefont {Kornrumpf}},\ and\ \bibinfo {author} {\bibfnamefont {D.}~\bibnamefont {Riegel}},\ }\bibfield  {title} {\bibinfo {title} {{Theory of the Temperature Dependence of the Electric Field Gradient in Noncubic Metals}},\ }\href {https://doi.org/10.1103/PhysRevLett.37.357} {\bibfield  {journal} {\bibinfo  {journal} {Phys. Rev. Lett.}\ }\textbf {\bibinfo {volume} {37}},\ \bibinfo {pages} {357} (\bibinfo {year} {1976})}\BibitemShut {NoStop}%
\bibitem [{\citenamefont {Jena}\ and\ \citenamefont {Rath}(1981)}]{jena1981}%
  \BibitemOpen
  \bibfield  {author} {\bibinfo {author} {\bibfnamefont {P.}~\bibnamefont {Jena}}\ and\ \bibinfo {author} {\bibfnamefont {J.}~\bibnamefont {Rath}},\ }\bibfield  {title} {\bibinfo {title} {{Ab initio calculation of the temperature dependence of the electric field gradient in Be}},\ }\href {https://doi.org/10.1103/PhysRevB.23.3823} {\bibfield  {journal} {\bibinfo  {journal} {Phys. Rev. B}\ }\textbf {\bibinfo {volume} {23}},\ \bibinfo {pages} {3823} (\bibinfo {year} {1981})}\BibitemShut {NoStop}%
\bibitem [{\citenamefont {Haas}(2024)}]{haas2024}%
  \BibitemOpen
  \bibfield  {author} {\bibinfo {author} {\bibfnamefont {H.}~\bibnamefont {Haas}},\ }\bibfield  {title} {\bibinfo {title} {{Temperature dependence of electric-field gradient in Zn and Cd: Replacing the ${T}^{3/2}$ law}},\ }\href {https://doi.org/10.1103/PhysRevB.109.064104} {\bibfield  {journal} {\bibinfo  {journal} {Phys. Rev. B}\ }\textbf {\bibinfo {volume} {109}},\ \bibinfo {pages} {064104} (\bibinfo {year} {2024})}\BibitemShut {NoStop}%
\bibitem [{\citenamefont {Laguta}\ \emph {et~al.}(2023)\citenamefont {Laguta}, \citenamefont {Zagorodniy}, \citenamefont {Kuzian}, \citenamefont {Kondakova}, \citenamefont {Chlan}, \citenamefont {\ifmmode \check{R}\else \v{R}\fi{}ezn\'{\i}\ifmmode~\check{c}\else \v{c}\fi{}ek}, \citenamefont {\ifmmode \check{S}\else \v{S}\fi{}t\ifmmode~\check{e}\else \v{e}\fi{}p\'ankov\'a}, \citenamefont {Bohdanov}, \citenamefont {Hlinka},\ and\ \citenamefont {Ramesh}}]{laguta2023}%
  \BibitemOpen
  \bibfield  {author} {\bibinfo {author} {\bibfnamefont {V.}~\bibnamefont {Laguta}}, \bibinfo {author} {\bibfnamefont {Y.~O.}\ \bibnamefont {Zagorodniy}}, \bibinfo {author} {\bibfnamefont {R.~O.}\ \bibnamefont {Kuzian}}, \bibinfo {author} {\bibfnamefont {I.~V.}\ \bibnamefont {Kondakova}}, \bibinfo {author} {\bibfnamefont {V.}~\bibnamefont {Chlan}}, \bibinfo {author} {\bibfnamefont {R.}~\bibnamefont {\ifmmode \check{R}\else \v{R}\fi{}ezn\'{\i}\ifmmode~\check{c}\else \v{c}\fi{}ek}}, \bibinfo {author} {\bibfnamefont {H.}~\bibnamefont {\ifmmode \check{S}\else \v{S}\fi{}t\ifmmode~\check{e}\else \v{e}\fi{}p\'ankov\'a}}, \bibinfo {author} {\bibfnamefont {D.}~\bibnamefont {Bohdanov}}, \bibinfo {author} {\bibfnamefont {J.}~\bibnamefont {Hlinka}},\ and\ \bibinfo {author} {\bibfnamefont {R.}~\bibnamefont {Ramesh}},\ }\bibfield  {title} {\bibinfo {title} {{Low-temperature ground state structure of $\mathrm{PbTi}{\mathrm{O}}_{3}$}},\ }\href {https://doi.org/10.1103/PhysRevB.107.104107} {\bibfield  {journal} {\bibinfo
  {journal} {Phys. Rev. B}\ }\textbf {\bibinfo {volume} {107}},\ \bibinfo {pages} {104107} (\bibinfo {year} {2023})}\BibitemShut {NoStop}%
\bibitem [{\citenamefont {{de Jong}}\ \emph {et~al.}(1998)\citenamefont {{de Jong}}, \citenamefont {Visscher},\ and\ \citenamefont {Nieuwpoort}}]{dejong1998}%
  \BibitemOpen
  \bibfield  {author} {\bibinfo {author} {\bibfnamefont {W.}~\bibnamefont {{de Jong}}}, \bibinfo {author} {\bibfnamefont {L.}~\bibnamefont {Visscher}},\ and\ \bibinfo {author} {\bibfnamefont {W.}~\bibnamefont {Nieuwpoort}},\ }\bibfield  {title} {\bibinfo {title} {{On the bonding and the electric field gradient of the uranyl ion}},\ }\href {https://doi.org/https://doi.org/10.1016/S0166-1280(98)00347-9} {\bibfield  {journal} {\bibinfo  {journal} {Journal of Molecular Structure: THEOCHEM}\ }\textbf {\bibinfo {volume} {458}},\ \bibinfo {pages} {41} (\bibinfo {year} {1998})}\BibitemShut {NoStop}%
\bibitem [{\citenamefont {Herzig}\ \emph {et~al.}(2008)\citenamefont {Herzig}, \citenamefont {Fojud}, \citenamefont {Żogał}, \citenamefont {Pietraszko}, \citenamefont {Dukhnenko}, \citenamefont {Jurga},\ and\ \citenamefont {Shitsevalova}}]{herzig2008}%
  \BibitemOpen
  \bibfield  {author} {\bibinfo {author} {\bibfnamefont {P.}~\bibnamefont {Herzig}}, \bibinfo {author} {\bibfnamefont {Z.}~\bibnamefont {Fojud}}, \bibinfo {author} {\bibfnamefont {O.~J.}\ \bibnamefont {Żogał}}, \bibinfo {author} {\bibfnamefont {A.}~\bibnamefont {Pietraszko}}, \bibinfo {author} {\bibfnamefont {A.}~\bibnamefont {Dukhnenko}}, \bibinfo {author} {\bibfnamefont {S.}~\bibnamefont {Jurga}},\ and\ \bibinfo {author} {\bibfnamefont {N.}~\bibnamefont {Shitsevalova}},\ }\bibfield  {title} {\bibinfo {title} {{Electric-field-gradient tensor and charge densities in LaB$_6$: $^{11}$B nuclear-magnetic-resonance single-crystal investigations and first-principles calculations}},\ }\href {https://doi.org/10.1063/1.2903150} {\bibfield  {journal} {\bibinfo  {journal} {Journal of Applied Physics}\ }\textbf {\bibinfo {volume} {103}},\ \bibinfo {pages} {083534} (\bibinfo {year} {2008})}\BibitemShut {NoStop}%
\bibitem [{\citenamefont {Polák}\ and\ \citenamefont {Fišer}(2003)}]{polak2003}%
  \BibitemOpen
  \bibfield  {author} {\bibinfo {author} {\bibfnamefont {R.}~\bibnamefont {Polák}}\ and\ \bibinfo {author} {\bibfnamefont {J.}~\bibnamefont {Fišer}},\ }\bibfield  {title} {\bibinfo {title} {{A CASSCF/icMRCI study of the electric field gradient in low-lying electronic states of N$_2^+/$N$_2$}},\ }\href {https://doi.org/https://doi.org/10.1016/S0301-0104(03)00138-1} {\bibfield  {journal} {\bibinfo  {journal} {Chemical Physics}\ }\textbf {\bibinfo {volume} {290}},\ \bibinfo {pages} {177} (\bibinfo {year} {2003})}\BibitemShut {NoStop}%
\bibitem [{\citenamefont {Ansari}\ \emph {et~al.}(2023)\citenamefont {Ansari}, \citenamefont {Sauer},\ and\ \citenamefont {Mazin}}]{ansari2023}%
  \BibitemOpen
  \bibfield  {author} {\bibinfo {author} {\bibfnamefont {J.~N.}\ \bibnamefont {Ansari}}, \bibinfo {author} {\bibfnamefont {K.~L.}\ \bibnamefont {Sauer}},\ and\ \bibinfo {author} {\bibfnamefont {I.~I.}\ \bibnamefont {Mazin}},\ }\bibfield  {title} {\bibinfo {title} {{Slow tail of nematic spin fluctuations in $\mathrm{Ba}({\mathrm{Fe}}_{1\ensuremath{-}x}{\mathrm{Co}}_{x}{)}_{2}{\mathrm{As}}_{2}$: Insight from nuclear magnetic resonance}},\ }\href {https://doi.org/10.1103/PhysRevB.108.064516} {\bibfield  {journal} {\bibinfo  {journal} {Phys. Rev. B}\ }\textbf {\bibinfo {volume} {108}},\ \bibinfo {pages} {064516} (\bibinfo {year} {2023})}\BibitemShut {NoStop}%
\bibitem [{\citenamefont {Sauer}\ \emph {et~al.}(2003)\citenamefont {Sauer}, \citenamefont {Suits}, \citenamefont {Garroway},\ and\ \citenamefont {Miller}}]{sauer2003}%
  \BibitemOpen
  \bibfield  {author} {\bibinfo {author} {\bibfnamefont {K.~L.}\ \bibnamefont {Sauer}}, \bibinfo {author} {\bibfnamefont {B.~H.}\ \bibnamefont {Suits}}, \bibinfo {author} {\bibfnamefont {A.~N.}\ \bibnamefont {Garroway}},\ and\ \bibinfo {author} {\bibfnamefont {J.~B.}\ \bibnamefont {Miller}},\ }\bibfield  {title} {\bibinfo {title} {{Secondary echoes in three-frequency nuclear quadrupole resonance of spin-1 nuclei}},\ }\href {https://doi.org/10.1063/1.1545442} {\bibfield  {journal} {\bibinfo  {journal} {The Journal of Chemical Physics}\ }\textbf {\bibinfo {volume} {118}},\ \bibinfo {pages} {5071} (\bibinfo {year} {2003})}\BibitemShut {NoStop}%
\bibitem [{\citenamefont {Suits}(2006)}]{suits2006}%
  \BibitemOpen
  \bibfield  {author} {\bibinfo {author} {\bibfnamefont {B.~H.}\ \bibnamefont {Suits}},\ }\bibinfo {title} {{NUCLEAR QUADRUPOLE RESONANCE SPECTROSCOPY}},\ in\ \href {https://doi.org/10.1007/0-387-37590-2_2} {\emph {\bibinfo {booktitle} {Handbook of Applied Solid State Spectroscopy}}},\ \bibinfo {editor} {edited by\ \bibinfo {editor} {\bibfnamefont {D.~R.}\ \bibnamefont {Vij}}}\ (\bibinfo  {publisher} {Springer US},\ \bibinfo {address} {Boston, MA},\ \bibinfo {year} {2006})\ pp.\ \bibinfo {pages} {65--96}\BibitemShut {NoStop}%
\bibitem [{\citenamefont {Hughes}\ \emph {et~al.}(1993)\citenamefont {Hughes}, \citenamefont {Austin},\ and\ \citenamefont {Swanson}}]{hughes1993}%
  \BibitemOpen
  \bibfield  {author} {\bibinfo {author} {\bibfnamefont {W.~C.}\ \bibnamefont {Hughes}}, \bibinfo {author} {\bibfnamefont {J.~C.}\ \bibnamefont {Austin}},\ and\ \bibinfo {author} {\bibfnamefont {M.~L.}\ \bibnamefont {Swanson}},\ }\bibfield  {title} {\bibinfo {title} {{Orientation of the electric‐field gradient arising from a vacancy in Hg$_{0.79}$Cd$_{0.21}$Te}},\ }\href {https://doi.org/10.1063/1.354331} {\bibfield  {journal} {\bibinfo  {journal} {Journal of Applied Physics}\ }\textbf {\bibinfo {volume} {74}},\ \bibinfo {pages} {4943} (\bibinfo {year} {1993})}\BibitemShut {NoStop}%
\bibitem [{\citenamefont {Kanert}\ \emph {et~al.}(1969)\citenamefont {Kanert}, \citenamefont {Kotzur},\ and\ \citenamefont {Mehring}}]{kanert1969}%
  \BibitemOpen
  \bibfield  {author} {\bibinfo {author} {\bibfnamefont {O.}~\bibnamefont {Kanert}}, \bibinfo {author} {\bibfnamefont {D.}~\bibnamefont {Kotzur}},\ and\ \bibinfo {author} {\bibfnamefont {M.}~\bibnamefont {Mehring}},\ }\bibfield  {title} {\bibinfo {title} {{Influence of Point Defects and Dislocations on the Line Shape of Nuclear Magnetic Resonance Signals}},\ }\href {https://doi.org/https://doi.org/10.1002/pssb.19690360130} {\bibfield  {journal} {\bibinfo  {journal} {physica status solidi (b)}\ }\textbf {\bibinfo {volume} {36}},\ \bibinfo {pages} {291} (\bibinfo {year} {1969})}\BibitemShut {NoStop}%
\bibitem [{\citenamefont {Cohen}\ and\ \citenamefont {Reif}(1957)}]{cohen1957}%
  \BibitemOpen
  \bibfield  {author} {\bibinfo {author} {\bibfnamefont {M.}~\bibnamefont {Cohen}}\ and\ \bibinfo {author} {\bibfnamefont {F.}~\bibnamefont {Reif}},\ }\bibfield  {title} {\bibinfo {title} {{Quadrupole Effects in Nuclear Magnetic Resonance Studies of Solids}},\ }in\ \href {https://doi.org/https://doi.org/10.1016/S0081-1947(08)60105-8} {\emph {\bibinfo {booktitle} {Solid State Physics}}},\ \bibinfo {series} {Solid State Physics}, Vol.~\bibinfo {volume} {5},\ \bibinfo {editor} {edited by\ \bibinfo {editor} {\bibfnamefont {F.}~\bibnamefont {SEITZ}}\ and\ \bibinfo {editor} {\bibfnamefont {D.}~\bibnamefont {TURNBULL}}}\ (\bibinfo  {publisher} {Academic Press},\ \bibinfo {year} {1957})\ pp.\ \bibinfo {pages} {321--438}\BibitemShut {NoStop}%
\bibitem [{\citenamefont {Han}\ \emph {et~al.}(1988)\citenamefont {Han}, \citenamefont {Timken},\ and\ \citenamefont {Oldfield}}]{han1988}%
  \BibitemOpen
  \bibfield  {author} {\bibinfo {author} {\bibfnamefont {O.~H.}\ \bibnamefont {Han}}, \bibinfo {author} {\bibfnamefont {H.~K.~C.}\ \bibnamefont {Timken}},\ and\ \bibinfo {author} {\bibfnamefont {E.}~\bibnamefont {Oldfield}},\ }\bibfield  {title} {\bibinfo {title} {{Solid‐state ‘‘magic‐angle’’ sample‐spinning nuclear magnetic resonance spectroscopic study of group III–V (13–15) semiconductors}},\ }\href {https://doi.org/10.1063/1.455418} {\bibfield  {journal} {\bibinfo  {journal} {The Journal of Chemical Physics}\ }\textbf {\bibinfo {volume} {89}},\ \bibinfo {pages} {6046} (\bibinfo {year} {1988})},\ \Eprint {https://arxiv.org/abs/https://pubs.aip.org/aip/jcp/article-pdf/89/10/6046/18973743/6046\_1\_online.pdf} {https://pubs.aip.org/aip/jcp/article-pdf/89/10/6046/18973743/6046\_1\_online.pdf} \BibitemShut {NoStop}%
\bibitem [{\citenamefont {Bowers}\ and\ \citenamefont {Mueller}(2005)}]{bowers2005}%
  \BibitemOpen
  \bibfield  {author} {\bibinfo {author} {\bibfnamefont {G.~M.}\ \bibnamefont {Bowers}}\ and\ \bibinfo {author} {\bibfnamefont {K.~T.}\ \bibnamefont {Mueller}},\ }\bibfield  {title} {\bibinfo {title} {{Electric field gradient distributions about strontium nuclei in cubic and octahedrally symmetric crystal systems}},\ }\href {https://doi.org/10.1103/PhysRevB.71.224112} {\bibfield  {journal} {\bibinfo  {journal} {Phys. Rev. B}\ }\textbf {\bibinfo {volume} {71}},\ \bibinfo {pages} {224112} (\bibinfo {year} {2005})}\BibitemShut {NoStop}%
\bibitem [{\citenamefont {Lany}\ \emph {et~al.}(2001)\citenamefont {Lany}, \citenamefont {Ostheimer}, \citenamefont {Wolf},\ and\ \citenamefont {Wichert}}]{lany1980}%
  \BibitemOpen
  \bibfield  {author} {\bibinfo {author} {\bibfnamefont {S.}~\bibnamefont {Lany}}, \bibinfo {author} {\bibfnamefont {V.}~\bibnamefont {Ostheimer}}, \bibinfo {author} {\bibfnamefont {H.}~\bibnamefont {Wolf}},\ and\ \bibinfo {author} {\bibfnamefont {T.}~\bibnamefont {Wichert}},\ }\bibfield  {title} {\bibinfo {title} {{Defect identification by means of electric field gradient calculation}},\ }\href {https://doi.org/https://doi.org/10.1016/S0921-4526(01)00853-5} {\bibfield  {journal} {\bibinfo  {journal} {Physica B: Condensed Matter}\ }\textbf {\bibinfo {volume} {308-310}},\ \bibinfo {pages} {980} (\bibinfo {year} {2001})},\ \bibinfo {note} {international Conference on Defects in Semiconductors}\BibitemShut {NoStop}%
\bibitem [{\citenamefont {Ansari}\ and\ \citenamefont {Sauer}(2024)}]{ansari2024}%
  \BibitemOpen
  \bibfield  {author} {\bibinfo {author} {\bibfnamefont {J.~N.}\ \bibnamefont {Ansari}}\ and\ \bibinfo {author} {\bibfnamefont {K.~L.}\ \bibnamefont {Sauer}},\ }\bibfield  {title} {\bibinfo {title} {{Spin-3/2 nuclear magnetic resonance: Exact analytical solutions for aligned systems and implications for probing Fe-based superconductors}},\ }\href {https://doi.org/10.1103/PhysRevB.110.214422} {\bibfield  {journal} {\bibinfo  {journal} {Phys. Rev. B}\ }\textbf {\bibinfo {volume} {110}},\ \bibinfo {pages} {214422} (\bibinfo {year} {2024})}\BibitemShut {NoStop}%
\bibitem [{\citenamefont {Fenta}\ \emph {et~al.}(2021)\citenamefont {Fenta}, \citenamefont {Amorim}, \citenamefont {Gon{\c{c}}alves}, \citenamefont {Fortunato}, \citenamefont {Barbosa}, \citenamefont {Cottenier}, \citenamefont {Correia}, \citenamefont {Pereira},\ and\ \citenamefont {Amaral}}]{fenta2021}%
  \BibitemOpen
  \bibfield  {author} {\bibinfo {author} {\bibfnamefont {A.~S.}\ \bibnamefont {Fenta}}, \bibinfo {author} {\bibfnamefont {C.~O.}\ \bibnamefont {Amorim}}, \bibinfo {author} {\bibfnamefont {J.~N.}\ \bibnamefont {Gon{\c{c}}alves}}, \bibinfo {author} {\bibfnamefont {N.}~\bibnamefont {Fortunato}}, \bibinfo {author} {\bibfnamefont {M.~B.}\ \bibnamefont {Barbosa}}, \bibinfo {author} {\bibfnamefont {S.}~\bibnamefont {Cottenier}}, \bibinfo {author} {\bibfnamefont {J.~G.}\ \bibnamefont {Correia}}, \bibinfo {author} {\bibfnamefont {L.~M.~C.}\ \bibnamefont {Pereira}},\ and\ \bibinfo {author} {\bibfnamefont {V.~S.}\ \bibnamefont {Amaral}},\ }\bibfield  {title} {\bibinfo {title} {{The electric field gradient as a signature of the binding and the local structure of adatoms on graphene}},\ }\href {https://doi.org/10.1007/s00339-021-04722-3} {\bibfield  {journal} {\bibinfo  {journal} {Applied Physics A}\ }\textbf {\bibinfo {volume} {127}},\ \bibinfo {pages} {573} (\bibinfo {year} {2021})}\BibitemShut {NoStop}%
\bibitem [{\citenamefont {Fujii}\ \emph {et~al.}(2024)\citenamefont {Fujii}, \citenamefont {Janson}, \citenamefont {Yasuoka}, \citenamefont {Rosner}, \citenamefont {Prots}, \citenamefont {Burkhardt}, \citenamefont {Schmidt},\ and\ \citenamefont {Baenitz}}]{fujii2024}%
  \BibitemOpen
  \bibfield  {author} {\bibinfo {author} {\bibfnamefont {T.}~\bibnamefont {Fujii}}, \bibinfo {author} {\bibfnamefont {O.}~\bibnamefont {Janson}}, \bibinfo {author} {\bibfnamefont {H.}~\bibnamefont {Yasuoka}}, \bibinfo {author} {\bibfnamefont {H.}~\bibnamefont {Rosner}}, \bibinfo {author} {\bibfnamefont {Y.}~\bibnamefont {Prots}}, \bibinfo {author} {\bibfnamefont {U.}~\bibnamefont {Burkhardt}}, \bibinfo {author} {\bibfnamefont {M.}~\bibnamefont {Schmidt}},\ and\ \bibinfo {author} {\bibfnamefont {M.}~\bibnamefont {Baenitz}},\ }\bibfield  {title} {\bibinfo {title} {{Experimental nuclear quadrupole resonance and computational study of the structurally refined topological semimetal ${\mathrm{TaSb}}_{2}$}},\ }\href {https://doi.org/10.1103/PhysRevB.109.035116} {\bibfield  {journal} {\bibinfo  {journal} {Phys. Rev. B}\ }\textbf {\bibinfo {volume} {109}},\ \bibinfo {pages} {035116} (\bibinfo {year} {2024})}\BibitemShut {NoStop}%
\bibitem [{\citenamefont {Bonf\`a}\ \emph {et~al.}(2022)\citenamefont {Bonf\`a}, \citenamefont {Frassineti}, \citenamefont {Wilkinson}, \citenamefont {Prando}, \citenamefont {Isah}, \citenamefont {Wang}, \citenamefont {Spina}, \citenamefont {Joseph}, \citenamefont {Mitrovi\ifmmode~\acute{c}\else \'{c}\fi{}}, \citenamefont {De~Renzi}, \citenamefont {Blundell},\ and\ \citenamefont {Sanna}}]{bonfa2022}%
  \BibitemOpen
  \bibfield  {author} {\bibinfo {author} {\bibfnamefont {P.}~\bibnamefont {Bonf\`a}}, \bibinfo {author} {\bibfnamefont {J.}~\bibnamefont {Frassineti}}, \bibinfo {author} {\bibfnamefont {J.~M.}\ \bibnamefont {Wilkinson}}, \bibinfo {author} {\bibfnamefont {G.}~\bibnamefont {Prando}}, \bibinfo {author} {\bibfnamefont {M.~M.}\ \bibnamefont {Isah}}, \bibinfo {author} {\bibfnamefont {C.}~\bibnamefont {Wang}}, \bibinfo {author} {\bibfnamefont {T.}~\bibnamefont {Spina}}, \bibinfo {author} {\bibfnamefont {B.}~\bibnamefont {Joseph}}, \bibinfo {author} {\bibfnamefont {V.~F.}\ \bibnamefont {Mitrovi\ifmmode~\acute{c}\else \'{c}\fi{}}}, \bibinfo {author} {\bibfnamefont {R.}~\bibnamefont {De~Renzi}}, \bibinfo {author} {\bibfnamefont {S.~J.}\ \bibnamefont {Blundell}},\ and\ \bibinfo {author} {\bibfnamefont {S.}~\bibnamefont {Sanna}},\ }\bibfield  {title} {\bibinfo {title} {{Entanglement between Muon and $I > \frac{1}{2}$ Nuclear Spins as a Probe of Charge Environment}},\ }\href
  {https://doi.org/10.1103/PhysRevLett.129.097205} {\bibfield  {journal} {\bibinfo  {journal} {Phys. Rev. Lett.}\ }\textbf {\bibinfo {volume} {129}},\ \bibinfo {pages} {097205} (\bibinfo {year} {2022})}\BibitemShut {NoStop}%
\bibitem [{\citenamefont {Rocha-Rodrigues}\ \emph {et~al.}(2020)\citenamefont {Rocha-Rodrigues}, \citenamefont {Santos}, \citenamefont {Miranda}, \citenamefont {Oliveira}, \citenamefont {Correia}, \citenamefont {Assali}, \citenamefont {Petrilli}, \citenamefont {Ara\'ujo},\ and\ \citenamefont {Lopes}}]{rodrigues2020}%
  \BibitemOpen
  \bibfield  {author} {\bibinfo {author} {\bibfnamefont {P.}~\bibnamefont {Rocha-Rodrigues}}, \bibinfo {author} {\bibfnamefont {S.~S.~M.}\ \bibnamefont {Santos}}, \bibinfo {author} {\bibfnamefont {I.~P.}\ \bibnamefont {Miranda}}, \bibinfo {author} {\bibfnamefont {G.~N.~P.}\ \bibnamefont {Oliveira}}, \bibinfo {author} {\bibfnamefont {J.~G.}\ \bibnamefont {Correia}}, \bibinfo {author} {\bibfnamefont {L.~V.~C.}\ \bibnamefont {Assali}}, \bibinfo {author} {\bibfnamefont {H.~M.}\ \bibnamefont {Petrilli}}, \bibinfo {author} {\bibfnamefont {J.~P.}\ \bibnamefont {Ara\'ujo}},\ and\ \bibinfo {author} {\bibfnamefont {A.~M.~L.}\ \bibnamefont {Lopes}},\ }\bibfield  {title} {\bibinfo {title} {{${\mathrm{Ca}}_{3}{\mathrm{Mn}}_{2}{\mathrm{O}}_{7}$ structural path unraveled by atomic-scale properties: A combined experimental and ab initio study}},\ }\href {https://doi.org/10.1103/PhysRevB.101.064103} {\bibfield  {journal} {\bibinfo  {journal} {Phys. Rev. B}\ }\textbf {\bibinfo {volume} {101}},\ \bibinfo {pages} {064103}
  (\bibinfo {year} {2020})}\BibitemShut {NoStop}%
\bibitem [{\citenamefont {Song}\ \emph {et~al.}(2022)\citenamefont {Song}, \citenamefont {Zheng}, \citenamefont {Yu}, \citenamefont {Li}, \citenamefont {Nie}, \citenamefont {Shan}, \citenamefont {Zhao}, \citenamefont {Li}, \citenamefont {Kang}, \citenamefont {Wu}, \citenamefont {Zhou}, \citenamefont {Sun}, \citenamefont {Liu}, \citenamefont {Luo}, \citenamefont {Wang}, \citenamefont {Ying}, \citenamefont {Wan}, \citenamefont {Wu},\ and\ \citenamefont {Chen}}]{song2022}%
  \BibitemOpen
  \bibfield  {author} {\bibinfo {author} {\bibfnamefont {D.}~\bibnamefont {Song}}, \bibinfo {author} {\bibfnamefont {L.}~\bibnamefont {Zheng}}, \bibinfo {author} {\bibfnamefont {F.}~\bibnamefont {Yu}}, \bibinfo {author} {\bibfnamefont {J.}~\bibnamefont {Li}}, \bibinfo {author} {\bibfnamefont {L.}~\bibnamefont {Nie}}, \bibinfo {author} {\bibfnamefont {M.}~\bibnamefont {Shan}}, \bibinfo {author} {\bibfnamefont {D.}~\bibnamefont {Zhao}}, \bibinfo {author} {\bibfnamefont {S.}~\bibnamefont {Li}}, \bibinfo {author} {\bibfnamefont {B.}~\bibnamefont {Kang}}, \bibinfo {author} {\bibfnamefont {Z.}~\bibnamefont {Wu}}, \bibinfo {author} {\bibfnamefont {Y.}~\bibnamefont {Zhou}}, \bibinfo {author} {\bibfnamefont {K.}~\bibnamefont {Sun}}, \bibinfo {author} {\bibfnamefont {K.}~\bibnamefont {Liu}}, \bibinfo {author} {\bibfnamefont {X.}~\bibnamefont {Luo}}, \bibinfo {author} {\bibfnamefont {Z.}~\bibnamefont {Wang}}, \bibinfo {author} {\bibfnamefont {J.}~\bibnamefont {Ying}}, \bibinfo {author} {\bibfnamefont {X.}~\bibnamefont
  {Wan}}, \bibinfo {author} {\bibfnamefont {T.}~\bibnamefont {Wu}},\ and\ \bibinfo {author} {\bibfnamefont {X.}~\bibnamefont {Chen}},\ }\bibfield  {title} {\bibinfo {title} {{Orbital ordering and fluctuations in a kagome superconductor CsV$_3$Sb$_5$}},\ }\href {https://doi.org/10.1007/s11433-021-1826-1} {\bibfield  {journal} {\bibinfo  {journal} {Science China Physics, Mechanics {\&} Astronomy}\ }\textbf {\bibinfo {volume} {65}},\ \bibinfo {pages} {247462} (\bibinfo {year} {2022})}\BibitemShut {NoStop}%
\bibitem [{\citenamefont {Banerjee}\ \emph {et~al.}(2021)\citenamefont {Banerjee}, \citenamefont {Sewak}, \citenamefont {Dey}, \citenamefont {Toprek},\ and\ \citenamefont {Pujari}}]{banerjee2021}%
  \BibitemOpen
  \bibfield  {author} {\bibinfo {author} {\bibfnamefont {D.}~\bibnamefont {Banerjee}}, \bibinfo {author} {\bibfnamefont {R.}~\bibnamefont {Sewak}}, \bibinfo {author} {\bibfnamefont {C.}~\bibnamefont {Dey}}, \bibinfo {author} {\bibfnamefont {D.}~\bibnamefont {Toprek}},\ and\ \bibinfo {author} {\bibfnamefont {P.}~\bibnamefont {Pujari}},\ }\bibfield  {title} {\bibinfo {title} {{Orthorhombic phases in bulk pure HfO$_2$: Experimental observation from perturbed angular correlation spectroscopy}},\ }\href {https://doi.org/https://doi.org/10.1016/j.mtcomm.2020.101827} {\bibfield  {journal} {\bibinfo  {journal} {Materials Today Communications}\ }\textbf {\bibinfo {volume} {26}},\ \bibinfo {pages} {101827} (\bibinfo {year} {2021})}\BibitemShut {NoStop}%
\bibitem [{\citenamefont {Das}\ and\ \citenamefont {Hahn}(1958)}]{das1958}%
  \BibitemOpen
  \bibfield  {author} {\bibinfo {author} {\bibfnamefont {T.~P.}\ \bibnamefont {Das}}\ and\ \bibinfo {author} {\bibfnamefont {E.~L.}\ \bibnamefont {Hahn}},\ }\href {https://catalog.hathitrust.org/Record/001482078} {\emph {\bibinfo {title} {{Nuclear quadrupole resonance spectroscopy}}}},\ Solid state physics. Supplement 1\ (\bibinfo  {publisher} {Academic Press},\ \bibinfo {address} {New York},\ \bibinfo {year} {1958})\BibitemShut {NoStop}%
\bibitem [{\citenamefont {Abragam}(1961)}]{abragam1961}%
  \BibitemOpen
  \bibfield  {author} {\bibinfo {author} {\bibfnamefont {A.}~\bibnamefont {Abragam}},\ }\href {https://global.oup.com/academic/product/principles-of-nuclear-magnetism-9780198520146} {\emph {\bibinfo {title} {{The principles of nuclear magnetism}}}}\ (\bibinfo  {publisher} {Oxford University Press},\ \bibinfo {year} {1961})\BibitemShut {NoStop}%
\bibitem [{\citenamefont {Zeldes}\ and\ \citenamefont {Livingston}(1957)}]{zeldes1957}%
  \BibitemOpen
  \bibfield  {author} {\bibinfo {author} {\bibfnamefont {H.}~\bibnamefont {Zeldes}}\ and\ \bibinfo {author} {\bibfnamefont {R.}~\bibnamefont {Livingston}},\ }\bibfield  {title} {\bibinfo {title} {{Zeeman Effect on the Quadrupole Spectra of Sodium, Potassium, and Barium Chlorates}},\ }\href {https://doi.org/10.1063/1.1743479} {\bibfield  {journal} {\bibinfo  {journal} {The Journal of Chemical Physics}\ }\textbf {\bibinfo {volume} {26}},\ \bibinfo {pages} {1102} (\bibinfo {year} {1957})}\BibitemShut {NoStop}%
\bibitem [{\citenamefont {GOLDMAN}(1990)}]{goldman1989}%
  \BibitemOpen
  \bibfield  {author} {\bibinfo {author} {\bibfnamefont {M.}~\bibnamefont {GOLDMAN}},\ }\bibfield  {title} {\bibinfo {title} {{Spin-$\frac{1}{2}$ Description of Spins $\frac{3}{2}$}},\ }in\ \href {https://doi.org/https://doi.org/10.1016/B978-0-12-025514-6.50008-3} {\emph {\bibinfo {booktitle} {Advances in Magnetic Resonance}}},\ \bibinfo {series} {Advances in Magnetic and Optical Resonance}, Vol.~\bibinfo {volume} {14},\ \bibinfo {editor} {edited by\ \bibinfo {editor} {\bibfnamefont {W.~S.}\ \bibnamefont {WARREN}}}\ (\bibinfo  {publisher} {Academic Press},\ \bibinfo {year} {1990})\ pp.\ \bibinfo {pages} {59--74}\BibitemShut {NoStop}%
\bibitem [{att()}]{attocube}%
  \BibitemOpen
  \href {https://www.attocube.com/en/products/nanopositioners/low-temperature-nanopositioners/anr240reslthv-rotator-360-endless} {\bibinfo {title} {{attocube ANR240/RES/LT/HV - rotator (360$\degree$ endless)}}}\BibitemShut {NoStop}%
\bibitem [{cry()}]{cryostat}%
  \BibitemOpen
  \href {https://www.montanainstruments.com/products/ca100r1} {\bibinfo {title} {{Montana Instruments $|$ CryoAdvance\textsuperscript{\textregistered} 100}}}\BibitemShut {NoStop}%
\bibitem [{\citenamefont {Nixon}\ and\ \citenamefont {Sauer}(2022)}]{nixon2022}%
  \BibitemOpen
  \bibfield  {author} {\bibinfo {author} {\bibfnamefont {K.~E.}\ \bibnamefont {Nixon}}\ and\ \bibinfo {author} {\bibfnamefont {K.~L.}\ \bibnamefont {Sauer}},\ }\bibfield  {title} {\bibinfo {title} {{Pulsed spin-locking of spin-3/2 nuclei: $^{39}$K-NQR of potassium chlorate}},\ }\href {https://doi.org/https://doi.org/10.1016/j.jmr.2022.107145} {\bibfield  {journal} {\bibinfo  {journal} {Journal of Magnetic Resonance}\ }\textbf {\bibinfo {volume} {335}},\ \bibinfo {pages} {107145} (\bibinfo {year} {2022})}\BibitemShut {NoStop}%
\bibitem [{\citenamefont {Xia}\ and\ \citenamefont {Ye}(1996)}]{xia1996}%
  \BibitemOpen
  \bibfield  {author} {\bibinfo {author} {\bibfnamefont {Y.}~\bibnamefont {Xia}}\ and\ \bibinfo {author} {\bibfnamefont {C.}~\bibnamefont {Ye}},\ }\bibfield  {title} {{\selectlanguage {english}\bibinfo {title} {{NQR spectroscopy of powder sample with spin I=1 and 3/2 (I) - Responses to pulses}}},\ }\href {https://experts.umn.edu/en/publications/nqr-spectroscopy-of-powder-sample-with-spin-i1-and-32-i-responses} {\bibfield  {journal} {\bibinfo  {journal} {Progress in Natural Science}\ }\textbf {\bibinfo {volume} {6}},\ \bibinfo {pages} {290} (\bibinfo {year} {1996})}\BibitemShut {NoStop}%
\bibitem [{\citenamefont {{Tecmag | A Magnetica Company}}(2024)}]{tecmag}%
  \BibitemOpen
  \bibfield  {author} {\bibinfo {author} {\bibnamefont {{Tecmag | A Magnetica Company}}},\ }\href {https://tecmag.com/} {\bibinfo {title} {{- TECMaG |}}} (\bibinfo {year} {2024})\BibitemShut {NoStop}%
\bibitem [{\citenamefont {Slade}()}]{slade2013}%
  \BibitemOpen
  \bibfield  {author} {\bibinfo {author} {\bibfnamefont {P.~G.}\ \bibnamefont {Slade}},\ }\bibinfo {title} {{The Arc and Interruption}},\ in\ \href {https://doi.org/10.1201/b15640} {\emph {\bibinfo {booktitle} {{Electrical Contacts: Principles and Applications, Second Edition}}}},\ \bibinfo {editor} {edited by\ \bibinfo {editor} {\bibfnamefont {P.~G.}\ \bibnamefont {Slade}}}\ (\bibinfo  {publisher} {CRC Press},\ \bibinfo {address} {Boca Raton})\ p.\ \bibinfo {pages} {592}\BibitemShut {NoStop}%
\bibitem [{\citenamefont {Holden}\ and\ \citenamefont {Morrison}(1982)}]{holden_crystals}%
  \BibitemOpen
  \bibfield  {author} {\bibinfo {author} {\bibfnamefont {A.}~\bibnamefont {Holden}}\ and\ \bibinfo {author} {\bibfnamefont {P.}~\bibnamefont {Morrison}},\ }\href {https://mitpress.mit.edu/9780262580502/crystals-and-crystal-growing/} {\emph {\bibinfo {title} {{Crystals and crystal growing}}}}\ (\bibinfo  {publisher} {MIT press},\ \bibinfo {year} {1982})\BibitemShut {NoStop}%
\bibitem [{\citenamefont {Teles}\ \emph {et~al.}(2015)\citenamefont {Teles}, \citenamefont {Rivera-Ascona}, \citenamefont {Polli}, \citenamefont {Oliveira-Silva}, \citenamefont {Vidoto}, \citenamefont {Andreeta},\ and\ \citenamefont {Bonagamba}}]{teles2015}%
  \BibitemOpen
  \bibfield  {author} {\bibinfo {author} {\bibfnamefont {J.}~\bibnamefont {Teles}}, \bibinfo {author} {\bibfnamefont {C.}~\bibnamefont {Rivera-Ascona}}, \bibinfo {author} {\bibfnamefont {R.~S.}\ \bibnamefont {Polli}}, \bibinfo {author} {\bibfnamefont {R.}~\bibnamefont {Oliveira-Silva}}, \bibinfo {author} {\bibfnamefont {E.~L.~G.}\ \bibnamefont {Vidoto}}, \bibinfo {author} {\bibfnamefont {J.~P.}\ \bibnamefont {Andreeta}},\ and\ \bibinfo {author} {\bibfnamefont {T.~J.}\ \bibnamefont {Bonagamba}},\ }\bibfield  {title} {\bibinfo {title} {{Experimental implementation of quantum information processing by Zeeman-perturbed nuclear quadrupole resonance}},\ }\href {https://doi.org/10.1007/s11128-015-0967-3} {\bibfield  {journal} {\bibinfo  {journal} {Quantum Information Processing}\ }\textbf {\bibinfo {volume} {14}},\ \bibinfo {pages} {1889} (\bibinfo {year} {2015})}\BibitemShut {NoStop}%
\bibitem [{\citenamefont {Odin}(1999)}]{odin1999}%
  \BibitemOpen
  \bibfield  {author} {\bibinfo {author} {\bibfnamefont {C.}~\bibnamefont {Odin}},\ }\bibfield  {title} {\bibinfo {title} {{Calculations of Multipulse Sequence in NQR of Spins $\frac{3}{2}$}},\ }\href {https://doi.org/https://doi.org/10.1006/jmre.1999.1905} {\bibfield  {journal} {\bibinfo  {journal} {Journal of Magnetic Resonance}\ }\textbf {\bibinfo {volume} {141}},\ \bibinfo {pages} {239} (\bibinfo {year} {1999})}\BibitemShut {NoStop}%
\bibitem [{\citenamefont {Choudhary}\ \emph {et~al.}(2020)\citenamefont {Choudhary}, \citenamefont {Ansari}, \citenamefont {Mazin},\ and\ \citenamefont {Sauer}}]{choudhary2020}%
  \BibitemOpen
  \bibfield  {author} {\bibinfo {author} {\bibfnamefont {K.}~\bibnamefont {Choudhary}}, \bibinfo {author} {\bibfnamefont {J.~N.}\ \bibnamefont {Ansari}}, \bibinfo {author} {\bibfnamefont {I.~I.}\ \bibnamefont {Mazin}},\ and\ \bibinfo {author} {\bibfnamefont {K.~L.}\ \bibnamefont {Sauer}},\ }\bibfield  {title} {\bibinfo {title} {{Density functional theory-based electric field gradient database}},\ }\href {https://doi.org/10.1038/s41597-020-00707-8} {\bibfield  {journal} {\bibinfo  {journal} {Scientific Data}\ }\textbf {\bibinfo {volume} {7}} (\bibinfo {year} {2020})}\BibitemShut {NoStop}%
\bibitem [{\citenamefont {Groth}(1908)}]{groth_german}%
  \BibitemOpen
  \bibfield  {author} {\bibinfo {author} {\bibfnamefont {P.}~\bibnamefont {Groth}},\ }\href {https://www.mindat.org/reference.php?id=17557086} {\emph {\bibinfo {title} {{Chemische Krystallographie}}}},\ Vol.~\bibinfo {volume} {2}\ (\bibinfo  {publisher} {Wilhem Engelmann},\ \bibinfo {year} {1908})\ pp.\ \bibinfo {pages} {90--92}\BibitemShut {NoStop}%
\bibitem [{\citenamefont {Schlick}(1972)}]{schlick1972}%
  \BibitemOpen
  \bibfield  {author} {\bibinfo {author} {\bibfnamefont {S.}~\bibnamefont {Schlick}},\ }\bibfield  {title} {\bibinfo {title} {{ESR Spectrum of O$_3^-$ Trapped in a Single Crystal of Potassium Chlorate}},\ }\href {https://doi.org/10.1063/1.1676919} {\bibfield  {journal} {\bibinfo  {journal} {The Journal of Chemical Physics}\ }\textbf {\bibinfo {volume} {56}},\ \bibinfo {pages} {654} (\bibinfo {year} {1972})}\BibitemShut {NoStop}%
\bibitem [{\citenamefont {Danielsen}\ \emph {et~al.}(1981)\citenamefont {Danielsen}, \citenamefont {Hazell},\ and\ \citenamefont {Larsen}}]{danielsen1981}%
  \BibitemOpen
  \bibfield  {author} {\bibinfo {author} {\bibfnamefont {J.}~\bibnamefont {Danielsen}}, \bibinfo {author} {\bibfnamefont {A.}~\bibnamefont {Hazell}},\ and\ \bibinfo {author} {\bibfnamefont {F.~K.}\ \bibnamefont {Larsen}},\ }\bibfield  {title} {\bibinfo {title} {{The structure of potassium chlorate at 77 and 298 K}},\ }\href {https://doi.org/10.1107/S0567740881004573} {\bibfield  {journal} {\bibinfo  {journal} {Acta Crystallographica Section B}\ }\textbf {\bibinfo {volume} {37}},\ \bibinfo {pages} {913} (\bibinfo {year} {1981})}\BibitemShut {NoStop}%
\bibitem [{\citenamefont {Utton}(1967)}]{utton1967}%
  \BibitemOpen
  \bibfield  {author} {\bibinfo {author} {\bibfnamefont {D.~B.}\ \bibnamefont {Utton}},\ }\bibfield  {title} {\bibinfo {title} {{Nuclear Quadrupole Resonance Thermometry}},\ }\href {https://doi.org/10.1088/0026-1394/3/4/002} {\bibfield  {journal} {\bibinfo  {journal} {Metrologia}\ }\textbf {\bibinfo {volume} {3}},\ \bibinfo {pages} {98} (\bibinfo {year} {1967})}\BibitemShut {NoStop}%
\bibitem [{\citenamefont {Weber}(1961)}]{weber1961}%
  \BibitemOpen
  \bibfield  {author} {\bibinfo {author} {\bibfnamefont {M.}~\bibnamefont {Weber}},\ }\bibfield  {title} {\bibinfo {title} {{Nuclear quadrupole spin-lattice relaxation in solids}},\ }\href {https://doi.org/https://doi.org/10.1016/0022-3697(61)90192-5} {\bibfield  {journal} {\bibinfo  {journal} {Journal of Physics and Chemistry of Solids}\ }\textbf {\bibinfo {volume} {17}},\ \bibinfo {pages} {267} (\bibinfo {year} {1961})}\BibitemShut {NoStop}%
\bibitem [{\citenamefont {Bayer}(1951)}]{bayer1951}%
  \BibitemOpen
  \bibfield  {author} {\bibinfo {author} {\bibfnamefont {H.}~\bibnamefont {Bayer}},\ }\bibfield  {title} {\bibinfo {title} {{Zur Theorie der Spin-Gitterrelaxation in Molekülkristallen}},\ }\href {https://doi.org/10.1007/bf01337696} {\bibfield  {journal} {\bibinfo  {journal} {The European Physical Journal A}\ }\textbf {\bibinfo {volume} {130}},\ \bibinfo {pages} {227} (\bibinfo {year} {1951})}\BibitemShut {NoStop}%
\bibitem [{\citenamefont {Ayant}(1956)}]{ayant1956}%
  \BibitemOpen
  \bibfield  {author} {\bibinfo {author} {\bibfnamefont {Y.}~\bibnamefont {Ayant}},\ }\bibfield  {title} {\bibinfo {title} {{La th{\'e}orie des temps de relaxation en r{\'e}sonance quadrupolaire}},\ }\href {https://doi.org/10.1051/jphysrad:01956001704033800} {\bibfield  {journal} {\bibinfo  {journal} {{Journal de Physique et le Radium}}\ }\textbf {\bibinfo {volume} {17}},\ \bibinfo {pages} {338} (\bibinfo {year} {1956})}\BibitemShut {NoStop}%
\bibitem [{\citenamefont {{Van Kranendonk}}(1954)}]{vankranendonk1954}%
  \BibitemOpen
  \bibfield  {author} {\bibinfo {author} {\bibfnamefont {J.}~\bibnamefont {{Van Kranendonk}}},\ }\bibfield  {title} {\bibinfo {title} {{Theory of quadrupolar nuclear spin-lattice relaxation}},\ }\href {https://doi.org/https://doi.org/10.1016/S0031-8914(54)80191-1} {\bibfield  {journal} {\bibinfo  {journal} {Physica}\ }\textbf {\bibinfo {volume} {20}},\ \bibinfo {pages} {781} (\bibinfo {year} {1954})}\BibitemShut {NoStop}%
\bibitem [{\citenamefont {Ramesh}\ \emph {et~al.}(2008)\citenamefont {Ramesh}, \citenamefont {Suresh}, \citenamefont {Raghavendra~Rao},\ and\ \citenamefont {Ramakrishna}}]{ramesh2008}%
  \BibitemOpen
  \bibfield  {author} {\bibinfo {author} {\bibfnamefont {K.~P.}\ \bibnamefont {Ramesh}}, \bibinfo {author} {\bibfnamefont {K.~S.}\ \bibnamefont {Suresh}}, \bibinfo {author} {\bibfnamefont {C.}~\bibnamefont {Raghavendra~Rao}},\ and\ \bibinfo {author} {\bibfnamefont {J.}~\bibnamefont {Ramakrishna}},\ }\bibfield  {title} {\bibinfo {title} {{Pressure and temperature dependence of the chlorine NQR in caesium and sodium chlorates}},\ }\href {https://doi.org/https://doi.org/10.1002/mrc.2206} {\bibfield  {journal} {\bibinfo  {journal} {Magnetic Resonance in Chemistry}\ }\textbf {\bibinfo {volume} {46}},\ \bibinfo {pages} {525} (\bibinfo {year} {2008})}\BibitemShut {NoStop}%
\bibitem [{\citenamefont {Woessner}\ and\ \citenamefont {Gutowsky}(1963)}]{woessner1963}%
  \BibitemOpen
  \bibfield  {author} {\bibinfo {author} {\bibfnamefont {D.~E.}\ \bibnamefont {Woessner}}\ and\ \bibinfo {author} {\bibfnamefont {H.~S.}\ \bibnamefont {Gutowsky}},\ }\bibfield  {title} {\bibinfo {title} {{Nuclear Pure Quadrupole Relaxation and Its Temperature Dependence in Solids}},\ }\href {https://doi.org/10.1063/1.1734268} {\bibfield  {journal} {\bibinfo  {journal} {The Journal of Chemical Physics}\ }\textbf {\bibinfo {volume} {39}},\ \bibinfo {pages} {440} (\bibinfo {year} {1963})}\BibitemShut {NoStop}%
\bibitem [{\citenamefont {Korabel’nikov}\ and\ \citenamefont {Zhuravlev}(2016)}]{korabelnikov2016}%
  \BibitemOpen
  \bibfield  {author} {\bibinfo {author} {\bibfnamefont {D.~V.}\ \bibnamefont {Korabel’nikov}}\ and\ \bibinfo {author} {\bibfnamefont {Y.~N.}\ \bibnamefont {Zhuravlev}},\ }\bibfield  {title} {\bibinfo {title} {{Ab initio investigations of the elastic properties of chlorates and perchlorates}},\ }\href {https://doi.org/10.1134/s1063783416060251} {\bibfield  {journal} {\bibinfo  {journal} {Physics of the Solid State}\ }\textbf {\bibinfo {volume} {58}},\ \bibinfo {pages} {1166} (\bibinfo {year} {2016})}\BibitemShut {NoStop}%
\bibitem [{\citenamefont {Weber}\ and\ \citenamefont {Hahn}(1960)}]{weber-hahn1960}%
  \BibitemOpen
  \bibfield  {author} {\bibinfo {author} {\bibfnamefont {M.~J.}\ \bibnamefont {Weber}}\ and\ \bibinfo {author} {\bibfnamefont {E.~L.}\ \bibnamefont {Hahn}},\ }\bibfield  {title} {\bibinfo {title} {{Selective Spin Excitation and Relaxation in Nuclear Quadrupole Resonance}},\ }\href {https://doi.org/10.1103/PhysRev.120.365} {\bibfield  {journal} {\bibinfo  {journal} {Phys. Rev.}\ }\textbf {\bibinfo {volume} {120}},\ \bibinfo {pages} {365} (\bibinfo {year} {1960})}\BibitemShut {NoStop}%
\bibitem [{\citenamefont {Melnick}\ and\ \citenamefont {Whitehead}(1982)}]{melnick1982}%
  \BibitemOpen
  \bibfield  {author} {\bibinfo {author} {\bibfnamefont {S.}~\bibnamefont {Melnick}}\ and\ \bibinfo {author} {\bibfnamefont {M.}~\bibnamefont {Whitehead}},\ }\bibfield  {title} {\bibinfo {title} {{Temperature dependence of the apparent spin-spin relaxation times, $T_2^*$ measured with a super-regenerative oscillator on $^{35}$Cl containing molecules}},\ }\href {https://doi.org/https://doi.org/10.1016/0022-2860(82)85153-3} {\bibfield  {journal} {\bibinfo  {journal} {Journal of Molecular Structure}\ }\textbf {\bibinfo {volume} {83}},\ \bibinfo {pages} {9} (\bibinfo {year} {1982})}\BibitemShut {NoStop}%
\bibitem [{\citenamefont {Malone}\ \emph {et~al.}(2011)\citenamefont {Malone}, \citenamefont {McGillvray},\ and\ \citenamefont {Sauer}}]{malone2011}%
  \BibitemOpen
  \bibfield  {author} {\bibinfo {author} {\bibfnamefont {M.~W.}\ \bibnamefont {Malone}}, \bibinfo {author} {\bibfnamefont {M.}~\bibnamefont {McGillvray}},\ and\ \bibinfo {author} {\bibfnamefont {K.~L.}\ \bibnamefont {Sauer}},\ }\bibfield  {title} {\bibinfo {title} {{Revealing dipolar coupling with NQR off-resonant pulsed spin locking}},\ }\href {https://doi.org/10.1103/PhysRevB.84.214430} {\bibfield  {journal} {\bibinfo  {journal} {Phys. Rev. B}\ }\textbf {\bibinfo {volume} {84}},\ \bibinfo {pages} {214430} (\bibinfo {year} {2011})}\BibitemShut {NoStop}%
\end{thebibliography}%

\end{document}